\RequirePackage{fix-cm}
\documentclass[twocolumn]{article}          
\usepackage{graphicx}
\usepackage{epsfig}
\usepackage{listings}
\usepackage{multirow}
\usepackage[round, sort&compress, authoryear]{natbib}

\begin{document}
\addtolength{\oddsidemargin}{-.4in}
\addtolength{\evensidemargin}{-.4in}
\addtolength{\textwidth}{0.6in}

\addtolength{\topmargin}{-1.2in}
\addtolength{\textheight}{2.0in}

\setlength{\columnsep}{0.6cm}

\title{Faster GPU Based Genetic Programming Using A Two
Dimensional Stack}



\author{Darren M. Chitty
\\Department of Computer Science,
\\University of Bristol, Merchant Venturers Bldg,
\\Woodland Road, BRISTOL BS8 1UB
\\email: darrenchitty@googlemail.com}



\maketitle

\begin{abstract}
Genetic Programming (GP) is a computationally intensive technique
which also has a high degree of natural parallelism. Parallel
computing architectures have become commonplace especially with
regards Graphics Processing Units (GPU). Hence, versions of GP
have been implemented that utilise these highly parallel computing
platforms enabling significant gains in the computational speed of
GP to be achieved. However, recently a two dimensional stack
approach to GP using a multi-core CPU also demonstrated
considerable performance gains. Indeed, performances equivalent to
or exceeding that achieved by a GPU were demonstrated. This paper
will demonstrate that a similar two dimensional stack approach can
also be applied to a GPU based approach to GP to better exploit
the underlying technology. Performance gains are achieved over a
standard single dimensional stack approach when utilising a GPU.
Overall, a peak computational speed of over 55 billion Genetic
Programming Operations per Second are observed, a two fold
improvement over the best GPU based single dimensional stack
approach from the literature.

\end{abstract}

\section{Introduction}
\label{sec:Introduction} Genetic Programming (GP)
\citep{Koza:1992} is an automated programming technique whereby
improved programs are generated using the principles of
evolutionary processes. GP can be considered a highly
computationally intensive algorithm which is mainly due to the
fact that traditionally an interpreter is used to evaluate
candidate GP programs. An interpreter is a slower methodology of
running a program due to a conditional statement being required at
each program step in order to establish which instruction should
be executed. Moreover, the candidate GP programs are often
re-interpreted over many fitness cases especially in the case of
classification or regression tasks. Additionally, GP is a
population based approach with traditionally a large number of
candidate GP programs making up the population.

Subsequently, given this high degree of computational complexity,
improving the execution speed of GP has been extensively studied.
Indeed, GP is naturally parallel through its population based
approach and its use of multiple fitness cases. As such,
considerable performance gains have been achieved by evaluating
both fitness cases and differing candidate GP programs
simultaneously by utilising parallel computing technology. With
parallel computing becoming ubiquitous with the development of
many-core Graphics Processing Units (GPUs) in desktop computers,
significant improvements in the execution speed of GP have been
achieved. However, a recent work has demonstrated equal
performance with GPU approaches to GP when using a multi-core CPU
\citep{Chitty:2012} even though GPUs have considerably greater
computational power. This performance was primarily achieved by
reducing the interpreter overhead by considering multiple fitness
cases at each step of a given interpreted program through the use
of a two dimensional stack. This approach also better exploited
fast cache memory which is also beneficial to improving execution
speed.

Given the success of this model, this paper will investigate
applying a two dimensional stack model for the purposes of GPU
based GP with the aim of improving the computational performance
by reducing the interpreter overheads and better exploitation of
fast cache memory. The paper is laid out as follows: Section 2
will introduce GPU technology and Section 3 will review GP
implementations that exploit modern parallel hardware. Section 4
will introduce the two dimensional stack model and the application
to a GPU based GP approach. Section 5 will introduce using a
linear GP representation which will considerably reduce stack
operations which will further benefit the two dimensional stack
model. Section 6 will consider further improvements in the use of
extended data types and the register file.

\section{Graphics Processing Unit Architecture}
\label{sec:GPUarch} Modern many-core GPUs have considerable
computing capability. This is achieved through the use of
thousands of Single Instruction Multiple Data (SIMD) processors
enabling massive parallelism to be achieved. Primarily, this
processing power is designed for the simulation of realistic
graphics required by modern computer games. However, this
technology has also been harnessed for high performance computing
needs with considerable gains having been realised for many
computationally intensive algorithms. There are two main
manufactures of GPUs, NVidia and AMD. This paper will focus on
NVidia GPUs specifically the Kepler version of the NVidia GPU, the
GK104.

The Kepler GPU consists of up to eight streaming multiprocessors
known as an SMX. Under each SMX there are 192 SIMD Stream
Processors (SP) which are restricted to executing the same
instruction simultaneously, a SIMD operation. However, SPs
operating under differing SMX can execute differing instructions
during the same clock cycle. This provides a limited level of
Multiple Instruction Multiple Data (MIMD) capability. Each SMX can
execute up to 2048 threads simultaneously. Threads are executed in
batches of 32 known as a \emph{warp}.

In terms of memory, the GK104 has both off-chip and on-chip
memory. Off-chip memory refers to memory which is located on the
GPU board. On-chip memory refers to memory that is located on each
SMX processor. There are two main off-chip memory types, global
memory and local memory. Global memory is accessible by all
threads executing under all SMX. This memory is stored in 32 banks
with consecutive memory addresses stored in differing banks. The
advantage of this is that if all 32 threads of a warp request a
memory location each of which is held in a differing bank, then
all the memory requests can be served faster using a single memory
fetch operation. However, if differing threads request a memory
location from the same bank this will take multiple memory fetch
operations which will slow performance. Thus for fast global
memory access, threads must access this global memory in a
contiguous manner. This means that a given thread does not access
consecutive memory locations unless all threads of a warp are
accessing the same memory location simultaneously. A second memory
area that is located off-chip is known as local memory. This is
private to each thread of execution and as such does not need to
be accessed in a contiguous manner.

On-chip memory is located on each SMX and is 64KB in size. As it
is located on-chip it has considerably faster access speeds than
off-chip memory. On-chip memory is used for both \emph{shared}
memory and \emph{L1 cache} memory. Shared memory can be accessed
by all threads executing under an SMX which as with global memory
must be accessed contiguously. L1 cache memory is not directly
addressable but is used to cache local memory accesses. Global
memory accesses are not cached in the L1 cache but in the off-chip
L2 cache. The size of shared and L1 cache memory can be configured
into three sizes. A preference for shared memory configures 48KB
of shared memory and 16KB for L1 cache. A preference for L1 cache
memory configures 48KB for the L1 cache and 16KB for the shared
memory. Finally, an equal preference configures each memory type
to have 32KB of storage.

\section{Accelerating Genetic Programming}
\label{sec:AccelGP} Genetic Programming (GP) \citep{Koza:1992} is
well known as being a computationally intensive technique through
the evaluation of thousands of candidate GP programs often over
large numbers of fitness cases. However, the technique is
naturally parallel with both candidate GP programs and fitness
cases being capable of being evaluated independently. Evaluating
candidate GP programs in parallel is known as a \emph{population
parallel} approach and evaluating fitness cases in parallel is
known as a \emph{data parallel} approach. In fact, GP can be
described as \emph{``embarrassingly parallel''} due to these two
differing degrees of parallelism. Subsequently, the technique is a
natural candidate for parallelisation in order to improve the
execution speed. An early implementation by Tufts
\citeyearpar{Tufts:1993} implemented a \emph{data parallel}
approach to GP whereby the fitness cases were evaluated in
parallel by multiple instances of candidate GP program executing
on a supercomputer. Juill\'e and Pollack \citeyearpar{Juille:1996}
implemented a Single Instruction Multiple Data (SIMD) model of GP
whereby multiple candidate solutions were evaluated in parallel,
using a MASPar MP-2 computer. Andre and Koza
\citeyearpar{Andre:1996} used a network of transputers with a
separate population of candidate solutions evolved on each
transputer with migration between the populations, an island
model. Niwa and Iba \citeyearpar{Niwa:1996} implemented a
distributed parallel implementation of GP, also using an island
model, executed over a Multiple Instruction Multiple Data (MIMD)
parallel system AP-1000 with 32 processors.

However, the first implementations of GP to harness the processing
power of GPUs were Chitty \citeyearpar{Chitty:2007} and Harding
and Banzhaf \citeyearpar{Harding:2007a}. Both compiled individual
candidate programs into a GPU kernel to be executed. Candidate GP
programs were evaluated in a sequential fashion on the GPU with
the parallelisation arising at the fitness case level, a
\emph{data parallel} approach. However, due to the length of time
it took to compile individual candidate programs, it was noted by
both papers that a large number of fitness cases were required to
offset the time penalty of performing the compilation step. Thus
significant gains in computational speed over a standard GP
implementation of up to 95 fold were demonstrated for problems
with large numbers of fitness cases. Langdon and Banzhaf
\citeyearpar{Langdon:2008:a} were the first to implement a more
traditional interpreter based approach to GP that operated on a
GPU. This approach exploited the fact that the interpreter could
operate in a SIMD manner with the same instruction being executed
at the same time on each processor when evaluating a single GP
program. However, the cost of executing an interpreter had an
impact on the speed and hence only a 7-12 fold performance gain
could be achieved over a CPU approach. Langdon and Banzhaf
\citeyearpar{Langdon:2008:a} also introduced a performance measure
known as Genetic Programming operations per second (GPop/s) to
demonstrate the speed of a given GP implementation. This is
calculated as the total number of nodes of all the GP trees
evaluated over the evolutionary process multiplied by the number
of fitness cases and divided by the total runtime of the GP
approach.

In 2007 NVidia released the Compute Unified Device Architecture
language (CUDA) specifically aimed at executing computationally
intensive tasks on a GPU. One of the first GP works to make use of
the CUDA language was that of Robilliard et al.
\citeyearpar{Robilliard:2008} who used the CUDA toolkit to
implement a traditional interpreter based approach. However, in
this case, fitness cases were evaluated in parallel (data
parallel) as were candidate programs (population parallel). This
is achieved by NVidia introducing up to eight multi-processors
(SMX) into their hardware which the SIMD processors (SPs) are
grouped under. The SIMD SP processors under a given SMX must
execute the same instruction at the same time. However, SP
processors grouped under one SMX can execute differing
instructions from those under another SMX. Subsequently, this
enables different candidate programs to be interpreted in parallel
by each multi-processor. The authors exploit this with an approach
known as \emph{BlockGP} whereby threads are separated into blocks
and each block executes under a differing SMX and evaluate a
differing candidate program with each thread in a block testing
against a differing fitness case. An eighty fold improvement over
a CPU based approach was demonstrated for a symbolic regression
problem. Robilliard et al. \citeyearpar{Robilliard:2009} later
demonstrate the advantage of the \emph{BlockGP} approach over an
implementation whereby differing threads of execution evaluate
differing candidate programs. This approach is known as
\emph{ThreadGP}. Moreover, an improvement to the \emph{BlockGP}
approach is used whereby the instructions of each candidate GP
program being evaluated by a block of threads is placed in shared
memory. Since threads will re-evaluate candidate programs for
multiple fitness cases, placing the candidate programs in shared
memory improves the access speed when fetching instructions. A
peak performance of 4 billion GPop/s is observed using this model.

Further CUDA approaches to GP have been considered. Lewis et al.
\citeyearpar{Lewis:2009} interleaved CPU operations with the
evaluations on a GPU to achieve a maximum performance gain of 3.8
billion GPop/s. Maitre et al. \citeyearpar{Maitre:2010} achieved
significant speedups even when using small numbers of fitness
cases by using efficient hardware scheduling in CUDA. A speedup of
up to 250x is achieved over a serial CPU implementation. Langdon
\citeyearpar{Langdon:2010} used a similar approach to
\emph{BlockGP} with the 37 multiplexer problem achieving a maximal
speed of 254 billion GPop/s through bit level parallelism. As the
problem under consideration is a boolean problem by using 32 bit
floats an extra 32x parallelism can be achieved. This is also
known as sub-machine-code GP \citep{Poli:1999aa}. Additionally, a
key difference in the work was that the stack was placed in the
faster shared memory rather than the slower local memory. Cano et
al. \citeyearpar{Cano:2012} implemented a contiguous GP evaluation
model evaluating three differing GP schemes using both a
multi-core CPU and two GPUs. A maximal performance gain of 820
fold is achieved using two GPUs over a serial Java CPU approach.
Cano et al. \citeyearpar{Cano:2013} further extend this work to
the evaluation of association rules achieving up to 67 billion
GPop/s when using two GPUs. Cano and Ventura
\citeyearpar{Cano:2014} also consider an additional level of
parallelism at the level of individual subtrees of a GP tree. If
two subtrees are independent of one another their output values
can be generated in parallel. A maximum speed of 22 billion GPop/s
was observed using this technique and a maximum speedup of 3.5x
over a standard approach to GPU based GP. Chitty
\citeyearpar{Chitty:2014} investigated extracting the best
performance from GPUs for the purposes of GP by exploiting the
limited faster memory resources to maximum effect. By using a
linear representation and registers for the lowest levels of the
stack a peak rate of 35 billion GPop/s was achieved. Augusto and
Barbosa \citeyearpar{Augusto:2013} utilise the alternative GPU
programming language, OpenCL. Both a data parallel and population
parallel approach is considered with a maximum speed of 13 billion
GPop/s achieved on a regression problem using the population
parallel approach.

Distributed versions have also been considered for the purposes of
parallel GP. Chong and Langdon \citeyearpar{Chong:1999} utilised
Java Servlets to distribute genetic programs across the Internet
whilst Klein and Spector \citeyearpar{Klein:2007} used Javascript
to harness computers connected to the Internet for the purposes of
GP without explicit user knowledge. Harding and Banzhaf
\citeyearpar{Harding:2009} devised a distributed GPU technique
whereby 14 computers equipped with graphics cards were utilised to
run GP. Each system handled part of the dataset and used a
compiled approach whereby a candidate solution was translated into
a CUDA kernel. A maximum speed 12.7 billion GPop/s was achieved
for the evolution of image filters. \cite{AlMadi:2013} used
MapReduce (Hadoop) to distribute candidate GP programs to an 18
node cluster which was found to be most suitable to large
population sizes. Finally, \cite{Sherry:2012} implemented an
island model GP system on the cloud using Amazon's EC2 reaching
350 nodes with an island on each node although a large number of
islands is required for any advantage to be realised.

\begin{table*}[ht]\centering
\footnotesize \centering \caption{A summary of the computation
speeds of approaches to parallel GP in terms of GPop/s (billions)
and speedup (in relation to a sequential implementation). In cases
whereby the information is not available N/A is used. Results
which use bit level parallelism are normalised.}
\begin{tabular}{|l|l|l|l|l|l|}
\hline \multirow{2}{*}\textbf{Author/s} &
\multirow{2}{*}\textbf{GP} & \multirow{2}{*}\textbf{Hardware} &
\multirow{2}{*}\textbf{Parallel} & \multirow{2}{*}\textbf{GPop/s}
& \multirow{2}{*}\textbf{Speedup}\\
& \textbf{Type} & & \textbf{Implementation} & & \\
\hline
\cite{Tufts:1993} & Tree GP & CM-5 & Data & N/A & N/A \\
\cite{Andre:1996} & Tree GP & 66x TRAMS Transputers & Island Population & N/A & linear+ \\
\cite{Juille:1996} & Canonical GP & MASPar MP-2 & Data & N/A & N/A \\
\cite{Niwa:1996} & Tree GP & AP-1000 & Island Population & N/A & N/A \\
\cite{Chong:1999} & Tree GP & N/A & Population & N/A & N/A \\
\cite{Martin:2001} & Linear GP & FPGA & Data & N/A & 8x \\
\cite{Eklund:2003} & Tree GP & FPGA & Diffusion Population & N/A & N/A \\
\cite{Chitty:2007} & Tree GP & NVidia 6400 GO & Data & N/A & 30x \\
\cite{Harding:2007a} & Cartesian GP & NVidia 7300 GO & Data & N/A & 95x \\
\cite{Klein:2007} & PUSH GP & N/A & Population & N/A & N/A \\
\cite{Langdon:2008:a} & Tree GP & NVidia 8800 GTX & Data & 1 & 12x \\
\cite{Langdon:2008:b} & Tree GP & NVidia 8800 GTX & Data & 0.5 & 7.6x \\
\cite{Robilliard:2008} & Tree GP & NVidia 8800 GTX & Data \& Population & N/A & 80x \\
\cite{Vasicek:2008} & Cartesian GP & FPGA & Data & N/A & 40x \\
\cite{Wilson:2008} & Linear GP & XBox 360 & Population & N/A & N/A \\
\cite{Harding:2009} & Cartesian GP & 14x NVidia 8200 & Data \& Population & 12.7 & N/A \\
\cite{Lewis:2009} & Cyclic & 2x NVidia 8800 GT, & Data \& Population & 3.8 & 435x \\
& Graph GP & 2x Intel Q6600 2.4GHz & & & \\
\cite{Robilliard:2009} & Tree GP & NVidia 8800 GTX & Data \& Population & 4 & 111x \\
\cite{Langdon:2010} & Tree GP & NVidia 295 GTX & Data \& Population & 21 & 313x \\
\cite{Maitre:2010} & Tree GP & 0.5x NVidia 295 GTX & Data \& Population & N/A & 250x \\
\cite{Wilson:2010} & Linear GP & XBox 360, & Population & N/A & 11x \\
& & NVidia 8800 GTX & & & \\
\cite{Cupertino:2011} & Linear GP & NVidia Tesla C1060 & Data \& Population & N/A & 5x \\
\cite{Lewis:2011} & TMBL GP & NVidia 260 GTX & Data & 200 & N/A \\
\cite{Cano:2012} & Tree GP & 2x NVidia 480 GTX & Data \& Population & N/A & 820x \\
\cite{Chitty:2012} & Tree GP & Intel i7 2600 & Data \& Population & 33 & 420x \\
\cite{Sherry:2012} & Tree GP & Amazon EC2 & Island Population & N/A & N/A \\
\cite{Vasicek:2012b} & Cartesian GP & Intel Core2 E8400 3GHz & N/A & N/A & 117x \\
\cite{AlMadi:2013} & Tree GP & 18x Intel Quad  & Population & N/A & 6x \\
& & Core CPU 2.67GHz & & & \\
\cite{Augusto:2013} & Tree GP & NVidia 285 GTX & Data \& Population & 13 & 126x \\
\cite{Cano:2013} & Tree GP & 2x NVidia 480 GTX & Data \& Population & 67 & 454x \\
\cite{Cano:2014} & Tree GP & NVidia 480 GTX & Data, Population \& & 22 & N/A \\
& & & Subtree & & \\
\cite{Chitty:2014} & Tree GP & NVidia 670 GTX & Data \& Population & 35 & N/A \\
\hline
\end{tabular} \centering
\label{tab:LitSummary}
\end{table*}

Some recent implementations of GP have considered a compiled
approach to candidate GP programs rather than an interpreted
approach. Lewis and Magoulas \citeyearpar{Lewis:2011} use CUDA to
compile kernels for GPU evaluation that represent candidate
programs. Moreover, they use a layer between the CUDA language and
a fully compiled CUDA kernel known as Parallel Thread EXecution
(PTX) to reduce the compilation time. An evaluation speed of 200
billion Tweaking Mutation Behaviour Learning (TMBL) operations per
second is achieved which can be interpreted as GPop/s. However,
this is just the evaluation phase of GP alone, the average
compilation time took approximately 0.05 seconds per individual
which significantly reduces the speed. \cite{Cupertino:2011} used
quantum inspired linear GP to generate PTX kernels to evaluate on
a GPU. Comparisons are made with a CPU version with a 25x speedup
reported for a large number of fitness cases. Comparisons are not
made with an interpreter based approach but it is noted that
compilation takes much longer than the evaluation time.
\cite{Vasicek:2012b} compiled candidate GP programs directly into
binary machine code for execution on a CPU. Speedups of up to 177x
were reported over an interpreted approach when using a single
precision floating point type. A compilation step can be avoided
by evolving machine code programs which can be executed directly
with no compilation or interpretation. \cite{Nordin:1994}
investigated the Automatic Induction of Machine Code with GP
(AIMGP) in order to speedup the process of evaluating candidate GP
programs. A speedup of 1500-2000x is reported over an equivalent
approach implemented in LISP. An average 60 fold speedup is
reported over an interpreted version implemented in C
\citep{Nordin:1995}.

Alternative parallel processor platforms have been considered for
the deployment of GP such as Microsoft's XBox 360
(\cite{Wilson:2008} and \cite{Wilson:2010}). Field Programable
Gate Arrays (FPGAs) have also been harnessed by the GP community
to accelerate GP. Martin \citeyearpar{Martin:2001} used a special
C compiler to run GP on FPGAs whilst Eklund
\citeyearpar{Eklund:2003} used an FPGA simulator to demonstrate
that GP could be run on FPGAs to model sun spot data. Vasicek and
Sekanina \citeyearpar{Vasicek:2008} used an FPGA with VPCs
(Virtual Reconfigurable Circuits) to evaluate candidate cartesian
GP programs. A speedup of 30-40 times over a basic CPU
implementation is reported.

\begin{figure*}[!ht]
\begin{center}
  \includegraphics[width=17cm]{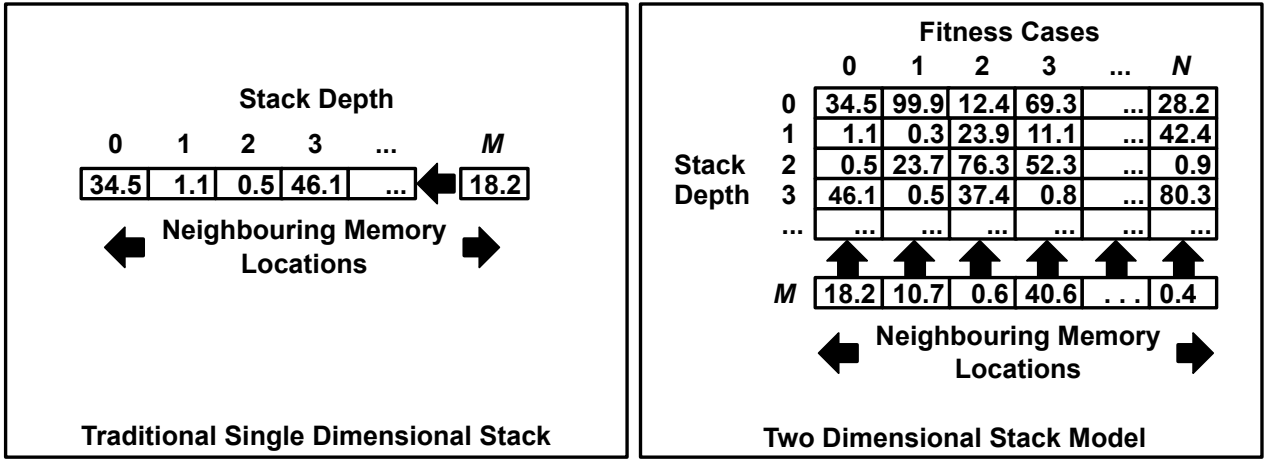}
\caption{The two differing stack models represented by a single
dimensional array and two a dimensional array in memory. With the
single dimensional stack, each level is a neighbouring memory
location of the last. With the two dimensional stack model, the
stack values for each fitness case need to be neighbouring memory
locations for optimal cache performance. Hence the stack needs to
be transposed for this to be the case.}
\label{fig:stackrep}       
\end{center}
\end{figure*}
\newpage
One final work of note is a parallel implementation of GP that
uses a multi-core CPU \citep{Chitty:2012}. The author implemented
a two-dimensional stack model whereby a single candidate GP
program instruction was evaluated over blocks of multiple fitness
cases rather than a single fitness case. This approach reduced the
interpreter overhead with reduced reinterpretation and facilitates
better exploitation of faster cache memory. Moreover, the approach
enabled the Streaming SIMD Extensions (SSE) instruction set to be
utilised boosting performance further. Additionally, parallel
threads of execution exploit the multiple cores and loop unrolling
exploits the Instruction Level Parallelism (ILP) of modern CPUs.
This approach enabled up to a 420x improvement in execution speed
over a basic serial implementation of GP. Moreover, a peak rate of
33 billion GPop/s is observed for a classification problem,
matching or exceeding current GPU implementations of GP. Table
\ref{tab:LitSummary} provides a comparison of these parallel GP
works specifying the hardware used and the speed achieved where
applicable.

\section{A Two Dimensional Stack Approach To GP}
\label{sec:2DGP} As mentioned in Section \ref{sec:AccelGP}, Chitty
\citeyearpar{Chitty:2012} introduced a modification to the
standard stack based approach to interpreted GP by implementing a
two dimensional stack to represent multiple fitness cases during a
single pass of the interpreter. This approach was shown to
considerably boost performance by reducing the number of times a
given program is re-interpreted and also better exploiting the
faster CPU cache memory. In fact performance was raised to such a
level that a parallel two dimensional stack CPU based approach to
GP could even match or outperform the best GPU based approaches to
GP from the literature.

With an interpreted approach to GP, a conditional statement is
required at each step of a candidate program to ascertain which
function to execute. In order to store outputs from functions and
also provide input values to a given function typically a stack is
used whereby required inputs are \emph{popped} off the top of the
stack and outputs from functions \emph{pushed} onto the top of the
stack. A standard single dimensional stack approach to GP operates
whereby a candidate GP program is interpreted such that each GP
function is executed using the data values held at the top of the
stack as inputs. The result is placed back on to the top of the
stack. Candidate GP programs are re-interpreted for each and every
fitness case which are used to measure the performance of the
program. This re-interpretation of candidate GP programs is
inefficient as a set of conditional statements need to be
evaluated at each step in order to ascertain which instruction to
execute. Moreover, each time a data value is retrieved from the
set of fitness cases in main memory, neighbouring memory locations
are placed into the faster cache memory. However, this data
contained within the cache memory is not properly utilised and the
extra performance that could be obtained is lost.

With the alternative two dimensional stack model, the first
dimension represents the levels of the stack and the second
dimension represents the number of fitness cases under
consideration in a single pass of the interpreter. Indeed, with a
two dimensional stack, the interpreter now operates in a SIMD
manner whereby whenever a function of a candidate GP program such
as addition is executed, it is performed across multiple fitness
cases. This significantly reduces the number of times that a
candidate GP program needs to be re-interpreted which will provide
a performance gain through greater efficiency. Secondly, a two
dimensional stack methodology ensures that cache memory is better
utilised as neighbouring memory locations within the stack will
now be accessed in a sequential manner. Using this approach with a
CPU and combining parallelism, Streaming SIMD Extensions (SSE) and
loop unrolling, speedups of over 420 fold over a single
dimensional, sequential stack implementation of GP were obtained
\citep{Chitty:2012}.

\begin{minipage}{0.45\textwidth}
\begin{lstlisting}[caption=A Two Dimensional Stack CUDA GP Interpreter,
  label=alg:2DBlockGP,basicstyle=\footnotesize,numberstyle=\footnotesize]
#define STACKDIM 4
__global__ void CUDAInterpreter(int* progs,
        float* inputs, float* outputs, float* results,
        int maxprogsize)
{
    extern __shared__ float prog[];
    for(int i=0;i<maxprogsize;i++) {
        prog[i]=progs[blockIdx.x*maxprogsize+i];
        if (prog[i]==255) break;
    }
    float stack[50][STACKDIM];float Sum=0.0;
    for(int i=threadIdx.x;i<NumCases;
        i+=NUM_THRDS*STACKDIM)
    {
        int ip=0; //instruction pointer
        int sp=0; //stack pointer
        while(prog[ip]!=255) {
            if (prog[ip]==CONST_VALUE) {
                for(int j=0;j<STACKDIM;j++)
                    stack[sp][j]=prog[ip];
                sp++;
            }
            else if (prog[ip]==DATA_INPUT) {
                for(j=0;j<STACKDIM;j++) {
                    int p=i+(j*NUM_THRDS);
                    stack[sp][j]=inputs[prog[ip]][p];
                }
                sp++;
            }
            else if (prog[ip]==ADDITION) {
                sp--;
                for(int j=0;j<STACKDIM;j++)
                    stack[sp][j]=stack[sp-1][j]+stack[sp][j];
            }
            else if (prog[ip]==SUBTRACTION) {
                sp--;
                for(int j=0;j<STACKDIM;j++)
                    stack[sp][j]=stack[sp-1][j]-stack[sp][j];
            }
            ip++;
        }
        for(int j=0;j<STACKDIM;j++)
            Sum+=GetError(stack[0][j],
                outputs[i+(j*NUM_THRDS)]);
    }
    results[blockIdx.x*NUM_THRDS+threadIdx.x]=Sum;
}
\end{lstlisting}
\end{minipage}

\subsection{Applying The Two Dimensional Stack Approach To A GPU
Implementation Of GP} \label{sec:2DGPGPU} Modern many-core GPUs
are significantly faster than their multi-core CPU counter parts.
Indeed, the NVidia GeForce Kepler 670 GTX graphics card has a
maximum speed of 1300 GFLOP/s whereas an Intel i7 2600 quad-core
CPU has a maximum speed of approximately 109 GFLOP/s. Thus on
paper, the GPU has over ten times the theoretical performance of
the CPU. However, these figures consider only the maximum
computing power and many algorithms are memory intensive and thus
cannot achieve this performance.

The current best GPU model of GP known as \emph{BlockGP}
\citep{Robilliard:2009} uses the SMX multi-processors of a GPU to
evaluate differing candidate GP programs in parallel whilst the
multiple threads of execution operating on the SPs under each SMX
operate on differing fitness cases in parallel. Candidate GP
program instructions are loaded into the fast shared memory for
quicker access whilst the stack is stored in local memory.
Accesses to this local memory will place stack memory locations
into the faster L1 cache whilst the fitness case data is stored in
global memory which gets cached in the off-chip L2 cache. Thus
making optimal use of the L1 cache involves using the stack
efficiently which a two dimensional stack approach should achieve.

As with the 2D stack approach for CPU-based GP, the stack held
within local memory on the GPU has two dimensions, the first being
the stack levels and the second representing the fitness cases
being considered within a single interpretation. Again, as with
the 2D stack approach for CPU-based GP, each GP instruction will
now be executed upon multiple fitness cases in succession. A key
difference with the GPU two dimensional stack implementation
presented here is how the fitness cases are accessed by each
thread of execution. With \emph{BlockGP}, each thread of execution
on the GPU evaluates a number of fitness cases equivalent to the
total number of fitness cases divided by the number of threads of
execution within a block. In order to obtain the maximal
performance from the global memory where fitness case data is
stored, each thread of execution does not evaluate consecutive
fitness cases thus ensuring faster contiguous memory access. This
is due to neighbouring memory locations being held in differing
memory banks such that for fast memory access, each thread of
execution should access a differing memory bank simultaneously.
However, with a two dimensional model, each time an input from the
fitness cases is required, multiple fitness case input values are
fetched from the memory. To ensure fast memory access, the
interpreter needs to fetch fitness cases from global memory
separated by the number of threads to ensure contiguous memory
access. Thus, if there are 32 threads of execution then each
interpreter thread needs to load the fitness case value as
determined by its thread identifier. Then the next fitness case
the interpreter will consider is the fitness case determined by
its thread identifier plus 32 and so forth. So when considering a
two dimensional stack model, the interpreter when accessing
fitness case values needs to place the fitness cases separated by
32 positions into consecutive local memory locations within the
two dimensional stack. A pseudo-code implementation of a two
dimensional approach to GP is shown in Listing
\ref{alg:2DBlockGP}.

The first aspect to note is that the loop defined on lines 12-13
only iterates over the number of fitness cases divided by the
number of threads of execution multiplied by the size of the
second dimension of the stack. Thus the number of iterations of
the interpreter are reduced by a factor relating to the size of
the two dimensional stack as defined by \emph{STACKDIM}. A further
aspect to note is that each GP operation consists of a loop
whereby consecutive memory locations of the stack are accessed at
the level as defined by the stack pointer \emph{sp}. This can be
observed on lines 32-33 whereby the GP function addition is
performed upon the top two levels of the 2D stack using a
\texttt{for} loop. Of final note is the method by which multiple
fitness case inputs are loaded onto the stack. This is shown on
lines 24-27. A pointer \emph{p} is used to select the appropriate
fitness case using the iterator from the main interpreter loop
plus the 2D stack iterator multiplied by the number of threads of
execution ensuring contiguous global memory accesses.

\subsection{Initial Results} \label{sec:InitResults}
In order to measure the performance of the two dimensional
approach to GP using a GPU, a comparison will be made with the
best traditional single dimensional approach implementation from
the literature known as \emph{BlockGP}. Four problem instances
will be considered, a symbolic regression problem, two
classification problems and the 20 multiplexor problem:
\\\\
\textbf{Sextic polynomial regression problem:} This problem is a
symbolic regression problem with the aim of establishing a
function which best approximates a set of data points. In this
case the set of data points is generated by the polynomial
$x^6-2x^4+x^2$ \citep{Koza:1992}. The function set used is {*, /,
+, -, $Sin$, $Cos$, $Log$, $Exp$}. The data point input values are
random values within the interval of [-1, +1]. For this problem
100000 fitness cases are used.
\\\\
\textbf{Shuttle classification problem:} This problem is a
classification problem whereby the goal is to establish a rule
that can correctly predict the class of a given set of input
values. In this case the data is the NASA Shuttle StatLog
\citep{king:1995} available from the machine learning repository
\citep{machinelearning:2010}. Both the test and training datasets
are used providing a total of 58000 fitness cases. There are 10
classes within the dataset with 75\% belonging to a single class.
There are nine input variables for each class. The function set
consists of {*, /, +, -, $>$, $<$, ==, AND, OR, IF} and the
terminal set consists of {X1...X9, -200.0 to 200.0}.
\\\\
\textbf{KDDcup intrusion classification problem:} This problem is
also a classification problem. In this case the dataset is the
test data from the KDD Cup 1999 problem \citeyearpar{kdd:1999}
with the aim to correctly classify potential network intrusions.
The dataset used is the test data which consists of 494021 fitness
cases with 41 input values. The function set consists of {+, -, *,
/, $>$, $<$, ==, AND, OR, IF, $Sin$, $Cos$, $Log$, $Exp$} and the
terminal set consists of {X1...X41, -20000.0 to 20000.0}. There
are 22 classes within the dataset.
\\\\
\textbf{Boolean 20-multiplexor problem:} This is a standard
Boolean problem used in GP \citep{Koza:1992}. The goal is to
establish a rule which takes address bits and data bits and
outputs the value of the data bit which the address bits specify.
The function set consists of {AND,OR,NAND,NOR} and the terminal
set consists of {A0-A3, D0-D15}. Every potential option is
considered such that there are $2^{20}$ fitness cases which is
1048576. However, in real terms this can be reduced as bit level
parallelism can be utilised such that each bit of a 32 bit
variable represents a differing fitness case, a technique also
known as sub-machine-code GP \citep{Poli:1999aa}. Thus, the number
of fitness cases are effectively reduced to 32768.

\begin{table}[!ht]
\footnotesize \centering \caption{GP parameters used throughout
paper}
\begin{tabular}{|c|c|}
\hline Population Size & 1000\\
\hline Max Generations & 50\\
\hline Max Tree Depth & 50\\
\hline Max Tree Size & 1000\\
\hline Selection & Tournament of size 7\\
\hline Crossover Probability & 0.95\\
\hline Mutation Probability & 0.2\\
\hline Regression Fitness & Mean squared error\\
\hline Classification Fitness & Sum of incorrect classifications\\
\hline
\end{tabular} \centering
\label{tab:params}
\end{table}

Results in this paper were generated using an NVidia GeForce
Kepler 670 GTX graphics card. Three memory configurations were
used for the division of shared memory and L1 cache memory to
establish the best setup for \emph{BlockGP}. A preference for L1
cache memory, a preference for shared memory or equal preferences.
For the two dimensional stack model, it is considered that a
preference for L1 cache is the best option as the technique favors
cache memory. A range of differing numbers of threads of execution
are used as are a range of sizes for the second dimension of the
2D stack model. The algorithms used were written in \texttt{C++}
and compiled using Microsoft Visual C++ 2010 and the GPU kernels
were compiled using NVidia's CUDA 5.0 toolkit. Table
\ref{tab:params} provides the GP parameters that were used
throughout this paper. These parameters are widely used within GP
but it should be noted that the goal of this research is improving
the speed of the evaluation phase of GP for which these GP
parameters only have a minimal effect. However, the best
methodology for comparing GP speeds is to perform a full GP run
for which these parameters are used. The results are averaged over
25 runs. Performance is described in terms of Genetic Programming
Operations per Second (GPop/s).

\begin{table*}[ht]\centering
\footnotesize \centering \caption{The GPop/s (measured in
billions) for each problem instance using the \emph{BlockGP}
single dimensional stack approach with differing preferences for
the on-chip memory and a range of threads within a block. The best
performance for each problem instance is shown in bold.}
\begin{tabular}{|c|c|c|c|c|}
\hline
\multirow{2}{*}\textbf{Problem} & \multirow{2}{*}\textbf{Num.} & \multicolumn{3}{c|}{\textbf{On-Chip Memory Preference}}\\
\cline{3-5} & \textbf{Threads} & Shared & Equal & L1\\
\hline & 128 & 23.010 $\pm$ 0.146 & 25.442 $\pm$ 0.094 & 26.184 $\pm$ 0.114\\
& 256 & 23.683 $\pm$ 0.202 & 25.969 $\pm$ 0.152 & \textbf{26.917} $\pm$ 0.124\\
& 384 & 23.092 $\pm$ 0.157 & 25.274 $\pm$ 0.142 & 26.047 $\pm$ 0.159\\
Sextic & 512 & 23.427 $\pm$ 0.156 & 25.699 $\pm$ 0.185 & 26.441 $\pm$ 0.132\\
& 640 & 22.749 $\pm$ 0.162 & 24.809 $\pm$ 0.135 & 25.365 $\pm$ 0.078\\
& 768 & 20.444 $\pm$ 0.103 & 21.983 $\pm$ 0.088 & 22.051 $\pm$ 0.047\\
& 896 & 21.777 $\pm$ 0.115 & 23.556 $\pm$ 0.082 & 23.795 $\pm$ 0.057\\
& 1024 & 22.578 $\pm$ 0.157 & 24.409 $\pm$ 0.132 & 24.847 $\pm$ 0.075\\

\hline & 128 & 21.161 $\pm$ 0.074 & 23.582 $\pm$ 0.105 & 24.925 $\pm$ 0.113\\
& 256 & 22.188 $\pm$ 0.063 & 24.864 $\pm$ 0.113 & \textbf{26.363} $\pm$ 0.099\\
& 384 & 21.766 $\pm$ 0.093 & 24.222 $\pm$ 0.126 & 25.512 $\pm$ 0.138\\
Shuttle & 512 & 22.163 $\pm$ 0.086 & 24.818 $\pm$ 0.119 & 26.204 $\pm$ 0.119\\
& 640 & 21.420 $\pm$ 0.071 & 23.778 $\pm$ 0.121 & 24.931 $\pm$ 0.114\\
& 768 & 18.953 $\pm$ 0.050 & 20.998 $\pm$ 0.053 & 21.288 $\pm$ 0.071\\
& 896 & 20.344 $\pm$ 0.052 & 22.465 $\pm$ 0.100 & 23.281 $\pm$ 0.076\\
& 1024 & 21.239 $\pm$ 0.083 & 23.657 $\pm$ 0.119 & 24.753 $\pm$ 0.100\\

\hline & 128 & 17.710 $\pm$ 0.014 & 22.952 $\pm$ 0.055 & 24.753 $\pm$ 0.062\\
& 256 & 18.738 $\pm$ 0.012 & 24.440 $\pm$ 0.054 & 26.458 $\pm$ 0.081\\
& 384 & 19.587 $\pm$ 0.012 & 24.112 $\pm$ 0.051 & 25.872 $\pm$ 0.072\\
KDDcup & 512 & 19.240 $\pm$ 0.008 & 25.062 $\pm$ 0.072 & 27.170 $\pm$ 0.076\\
& 640 & 19.692 $\pm$ 0.008 & 24.354 $\pm$ 0.055 & 26.128 $\pm$ 0.071\\
& 768 & 18.736 $\pm$ 0.003 & 21.775 $\pm$ 0.022 & 22.181 $\pm$ 0.003\\
& 896 & 20.060 $\pm$ 0.004 & 23.625 $\pm$ 0.007 & 24.973 $\pm$ 0.008\\
& 1024 & 19.343 $\pm$ 0.006 & 25.219 $\pm$ 0.069 & \textbf{27.255} $\pm$ 0.073\\

\hline & 128 & 632.803 $\pm$ 1.511 & 707.501 $\pm$ 1.741 & 596.036 $\pm$ 1.566\\
& 256 & 683.493 $\pm$ 1.378 & 784.310 $\pm$ 1.728 & 821.266 $\pm$ 2.014\\
& 384 & 701.902 $\pm$ 1.623 & 780.487 $\pm$ 2.288 & 837.306 $\pm$ 2.617\\
20-Mult. & 512 & 704.146 $\pm$ 2.007 & 805.312 $\pm$ 2.296 & \textbf{865.255} $\pm$ 2.501\\
& 640 & 700.425 $\pm$ 1.577 & 777.336 $\pm$ 2.312 & 831.874 $\pm$ 2.139\\
& 768 & 635.508 $\pm$ 1.070 & 696.895 $\pm$ 2.037 & 739.726 $\pm$ 1.588\\
& 896 & 678.915 $\pm$ 1.899 & 745.213 $\pm$ 2.324 & 794.778 $\pm$ 1.621\\
& 1024 & 688.996 $\pm$ 1.756 & 785.781 $\pm$ 2.192 & 840.027 $\pm$ 2.231\\
\hline
\end{tabular} \centering
\label{tab:RobilliardDynamic}
\end{table*}

\begin{table*}[ht]\centering
\footnotesize \centering \caption{The GPop/s (measured in
billions) for each problem instance using a two-dimensional stack
approach and a preference for greater L1 cache memory for a range
of threads within a block and a range of second dimension sizes.
The best performance for each problem instance is shown in bold.}
\begin{tabular}{|c|c|c|c|c|c|c|}
\hline
\multirow{2}{*}\textbf{Problem} & \multirow{2}{*}\textbf{Num.} & \multicolumn{5}{c|}{\textbf{Size of Second Dimension Of Stack}}\\
\cline{3-7} & \textbf{Threads} & 2 & 3 & 4 & 5 & 6\\
\hline & 32 & 13.538 $\pm$ 0.011 & 16.386 $\pm$ 0.037 & 18.930 $\pm$ 0.046 & 18.867 $\pm$ 0.050 & 19.680 $\pm$ 0.044\\
& 64 & 24.812 $\pm$ 0.142 & 27.635 $\pm$ 0.116 & 29.290 $\pm$ 0.155 & 27.769 $\pm$ 0.108 & 23.593 $\pm$ 0.112\\
& 96 & 31.192 $\pm$ 0.144 & 31.598 $\pm$ 0.208 & 28.971 $\pm$ 0.144 & 24.686 $\pm$ 0.071 & 20.506 $\pm$ 0.046\\
& 128 & 33.769 $\pm$ 0.217 & 28.471 $\pm$ 0.131 & 22.460 $\pm$ 0.056 & 20.294 $\pm$ 0.027 & 18.462 $\pm$ 0.031\\
& 256 & \textbf{34.205} $\pm$ 0.265 & 29.406 $\pm$ 0.097 & 22.697 $\pm$ 0.058 & 21.047 $\pm$ 0.025 & 18.593 $\pm$ 0.018\\
Sextic & 384 & 33.605 $\pm$ 0.207 & 30.622 $\pm$ 0.150 & 24.602 $\pm$ 0.055 & 21.508 $\pm$ 0.047 & 19.179 $\pm$ 0.037\\
& 512 & 33.398 $\pm$ 0.202 & 29.744 $\pm$ 0.113 & 23.490 $\pm$ 0.051 & 21.684 $\pm$ 0.031 & 18.545 $\pm$ 0.021\\
& 640 & 32.603 $\pm$ 0.231 & 29.615 $\pm$ 0.143 & 24.062 $\pm$ 0.076 & 21.178 $\pm$ 0.035 & 18.835 $\pm$ 0.048\\
& 768 & 29.892 $\pm$ 0.151 & 29.763 $\pm$ 0.209 & 27.969 $\pm$ 0.150 & 25.219 $\pm$ 0.125 & 20.270 $\pm$ 0.030\\
& 896 & 30.802 $\pm$ 0.162 & 29.133 $\pm$ 0.162 & 24.911 $\pm$ 0.065 & 21.676 $\pm$ 0.090 & 18.713 $\pm$ 0.049\\
& 1024 & 31.241 $\pm$ 0.203 & 27.442 $\pm$ 0.154 & 22.399 $\pm$ 0.042 & 20.541 $\pm$ 0.023 & 17.953 $\pm$ 0.019\\

\hline & 32 & 13.119 $\pm$ 0.015 & 17.264 $\pm$ 0.068 & 20.441 $\pm$ 0.082 & 20.812 $\pm$ 0.075 & 22.273 $\pm$ 0.060\\
& 64 & 23.862 $\pm$ 0.101 & 28.126 $\pm$ 0.129 & 28.085 $\pm$ 0.082 & 24.170 $\pm$ 0.033 & 21.844 $\pm$ 0.018\\
& 96 & 29.719 $\pm$ 0.108 & 26.686 $\pm$ 0.043 & 21.416 $\pm$ 0.027 & 18.873 $\pm$ 0.019 & 17.439 $\pm$ 0.014\\
& 128 & 27.933 $\pm$ 0.041 & 21.474 $\pm$ 0.024 & 18.166 $\pm$ 0.015 & 16.362 $\pm$ 0.014 & 15.431 $\pm$ 0.015\\
& 256 & 29.055 $\pm$ 0.047 & 21.940 $\pm$ 0.014 & 18.277 $\pm$ 0.013 & 16.319 $\pm$ 0.014 & 15.393 $\pm$ 0.009\\
Shuttle & 384 & \textbf{31.168} $\pm$ 0.100 & 23.712 $\pm$ 0.027 & 19.293 $\pm$ 0.016 & 16.920 $\pm$ 0.026 & 15.621 $\pm$ 0.010\\
& 512 & 29.159 $\pm$ 0.055 & 22.549 $\pm$ 0.030 & 18.215 $\pm$ 0.028 & 16.540 $\pm$ 0.012 & 15.524 $\pm$ 0.009\\
& 640 & 30.088 $\pm$ 0.118 & 22.866 $\pm$ 0.017 & 18.855 $\pm$ 0.014 & 16.663 $\pm$ 0.020 & 15.466 $\pm$ 0.011\\
& 768 & 29.605 $\pm$ 0.176 & 27.146 $\pm$ 0.043 & 22.092 $\pm$ 0.023 & 19.104 $\pm$ 0.022 & 17.184 $\pm$ 0.016\\
& 896 & 29.813 $\pm$ 0.106 & 23.394 $\pm$ 0.022 & 19.002 $\pm$ 0.015 & 17.031 $\pm$ 0.017 & 15.696 $\pm$ 0.013\\
& 1024 & 27.332 $\pm$ 0.080 & 21.458 $\pm$ 0.024 & 17.608 $\pm$ 0.017 & 15.646 $\pm$ 0.013 & 14.745 $\pm$ 0.014\\

\hline & 32 & 12.852 $\pm$ 0.002 & 17.001 $\pm$ 0.005 & 20.026 $\pm$ 0.010 & 20.108 $\pm$ 0.008 & 20.552 $\pm$ 0.010\\
& 64 & 23.939 $\pm$ 0.008 & 27.235 $\pm$ 0.026 & 21.661 $\pm$ 0.014 & 18.920 $\pm$ 0.006 & 14.849 $\pm$ 0.007\\
& 96 & 27.891 $\pm$ 0.024 & 20.456 $\pm$ 0.010 & 17.073 $\pm$ 0.010 & 15.378 $\pm$ 0.009 & 13.987 $\pm$ 0.005\\
& 128 & 21.235 $\pm$ 0.015 & 17.060 $\pm$ 0.009 & 15.077 $\pm$ 0.008 & 13.839 $\pm$ 0.007 & 13.557 $\pm$ 0.004\\
& 256 & 22.014 $\pm$ 0.006 & 17.526 $\pm$ 0.010 & 15.184 $\pm$ 0.007 & 13.978 $\pm$ 0.004 & 13.741 $\pm$ 0.004\\
KDDcup & 384 & 24.600 $\pm$ 0.006 & 18.678 $\pm$ 0.005 & 15.967 $\pm$ 0.006 & 14.417 $\pm$ 0.004 & 14.149 $\pm$ 0.002\\
& 512 & 22.902 $\pm$ 0.012 & 18.425 $\pm$ 0.004 & 15.526 $\pm$ 0.005 & 14.271 $\pm$ 0.002 & 14.446 $\pm$ 0.002\\
& 640 & 24.496 $\pm$ 0.023 & 18.511 $\pm$ 0.005 & 15.876 $\pm$ 0.004 & 14.430 $\pm$ 0.001 & 14.500 $\pm$ 0.016\\
& 768 & \textbf{30.536} $\pm$ 0.013 & 22.203 $\pm$ 0.005 & 18.373 $\pm$ 0.006 & 16.198 $\pm$ 0.003 & 13.845 $\pm$ 0.005\\
& 896 & 26.050 $\pm$ 0.008 & 19.351 $\pm$ 0.003 & 16.369 $\pm$ 0.002 & 14.843 $\pm$ 0.006 & 15.086 $\pm$ 0.002\\
& 1024 & 22.640 $\pm$ 0.007 & 18.337 $\pm$ 0.004 & 15.504 $\pm$ 0.003 & 14.277 $\pm$ 0.001 & 14.543 $\pm$ 0.001\\

\hline & 32 & 294.742 $\pm$ 0.302 & 368.967 $\pm$ 0.716 & 444.253 $\pm$ 0.456 & 427.078 $\pm$ 0.882 & 468.940 $\pm$ 0.659\\
& 64 & 536.480 $\pm$ 1.240 & 615.354 $\pm$ 1.335 & 699.182 $\pm$ 1.968 & 653.330 $\pm$ 1.765 & 669.642 $\pm$ 1.519\\
& 96 & 697.700 $\pm$ 1.701 & 730.826 $\pm$ 1.490 & 744.960 $\pm$ 0.750 & 677.033 $\pm$ 0.475 & 653.005 $\pm$ 0.356\\
& 128 & 784.397 $\pm$ 1.877 & 738.643 $\pm$ 0.966 & 685.319 $\pm$ 0.604 & 610.788 $\pm$ 0.387 & 580.573 $\pm$ 0.341\\
& 256 & 792.898 $\pm$ 1.547 & 642.027 $\pm$ 0.644 & 567.665 $\pm$ 0.435 & 515.719 $\pm$ 0.464 & 485.477 $\pm$ 0.331\\
20-Mult. & 384 & 803.756 $\pm$ 1.958 & 645.628 $\pm$ 0.625 & 565.830 $\pm$ 0.406 & 518.172 $\pm$ 0.390 & 482.823 $\pm$ 0.351\\
& 512 & 792.547 $\pm$ 1.445 & 636.935 $\pm$ 0.652 & 555.781 $\pm$ 0.541 & 516.372 $\pm$ 0.387 & 480.714 $\pm$ 0.228\\
& 640 & 792.899 $\pm$ 2.119 & 647.745 $\pm$ 1.187 & 562.651 $\pm$ 0.478 & 495.505 $\pm$ 0.280 & 478.474 $\pm$ 0.334\\
& 768 & \textbf{866.365} $\pm$ 2.623 & 727.432 $\pm$ 1.727 & 631.289 $\pm$ 0.868 & 560.068 $\pm$ 0.407 & 520.952 $\pm$ 0.439\\
& 896 & 815.287 $\pm$ 2.381 & 660.618 $\pm$ 0.809 & 573.290 $\pm$ 0.489 & 502.213 $\pm$ 0.476 & 487.595 $\pm$ 0.191\\
& 1024 & 771.872 $\pm$ 1.793 & 620.390 $\pm$ 0.816 & 547.441 $\pm$ 0.271 & 485.368 $\pm$ 0.326 & 451.957 $\pm$ 0.219\\
\hline
\end{tabular} \centering
\label{tab:2DstackGPU}
\end{table*}

To begin with, a baseline is established using the current best
GPU GP approach \emph{BlockGP} for a range of thread numbers and
three shared memory preferences. These results are shown in Table
\ref{tab:RobilliardDynamic}. The first point that can be made is
that it can be clearly seen that for the single dimensional stack
\emph{BlockGP} approach, a preference for a greater level of L1
cache memory rather than shared memory provides the best
performance for all problem instances. The key reason for this is
that very little shared memory is required to hold candidate GP
programs. However, the amount of local memory used to hold the
stacks for each thread of execution within a block is relatively
large and thus the L1 cache can greatly benefit the access speed
of the stack. A peak rate of 27.3 billion GPop/s are observed for
the KDDcup classification problem with 865 billion GPop/s for the
multiplexer problem which benefits from an additional 32x bitwise
parallelism due to its boolean nature.

Table \ref{tab:2DstackGPU} shows the results from using a two
dimensional stack implementation for a GPU. From these results it
can be observed that an improved performance can be achieved by
using a two dimensional stack approach. For the first three
problem instances, performances in excess of 30 billion GPop/s are
achieved with 864 billion GPop/s for the multiplexer problem. The
gain in computational speed over the the \emph{BlockGP} approach
is on average 14\% although the performance between the two
approaches is very similar for the multiplexer problem. The best
performance was observed for the sextic regression problem with 34
billion GPop/s, a 27\% performance gain over using a single
dimensional stack model. However, it should be noted that the best
performance from using a two dimensional model occurs when
considering only two fitness cases simultaneously, a size of two
for the second dimension of the stack. Performance degrades
sharply for larger considerations of fitness cases. This is in
stark contrast to the CPU version of a two dimensional stack
whereby the size of the second dimension of the stack and hence
the number of fitness cases that can be considered simultaneously
was in the order of thousands \citep{Chitty:2012}. Clearly,
increasing the size of the second dimension of the stack reduces
the level of re-interpretation required which should result in a
performance gain. However, instead a performance deficit is
observed which indicates that the cache memory cannot accommodate
the extra stack requirements of a larger second dimension size to
the extent that this more than offsets the performance advantage
gained from reduced re-interpretation of candidate GP programs.
\newpage
This theory is reinforced by observing that as the size of the
second dimension of the 2D stack increases, the number of threads
of execution which provide the best performance decreases. Indeed,
when considering a second dimension size of six, for two of the
problem instances the best performance arises when using only 32
threads of execution resulting in most of the SPs under an SMX
being idle. The amount of L1 cache memory available in each SMX is
much smaller than that of a CPU. So as the number of threads of
execution under each SMX multi-processor increases, the pressure
on the L1 cache is increased through the use of more local memory
being needed to hold the local stacks of each thread of execution.
However, to obtain a high level of performance for a GPU, a large
number of threads are required as there are hundreds of SPs under
each SMX. Thus, there are two sources of pressure on the L1 cache
memory, the size of the second dimension of the 2D stack model and
the number of threads of execution. Consequently, a balance must
be struck between the benefits of efficient interpretation of
candidate GP programs using a 2D stack and full utilisation of the
SPs under each SMX.

\begin{table*}[ht]\centering
\footnotesize \centering \caption{The percentages of candidate GP
programs from the experiments conducted in this paper that can be
evaluated with a given stack size limit and an RPN representation}
\begin{tabular}{|c|c|c|c|c|c|c|c|c|c|}
\hline
\multirow{2}{*}\textbf{Problem} & \multicolumn{9}{c|}{\textbf{Stack Limit}}\\
\cline{2-10} & 2 & 3 & 4 & 5 & 6 & 7 & 8 & 9 & 10\\
\hline Sextic & 18.71 & 35.47 & 84.72 & 95.37 & 98.62 & 99.78 & 99.94 & 99.98 & 100.0\\
Shuttle & 3.91 & 17.89 & 77.18 & 91.00 & 95.97 & 98.40 & 99.26 & 99.59 & 99.86\\
KDDcup & 5.54 & 36.16 & 79.97 & 93.73 & 97.73 & 99.05 & 99.68 & 99.91 & 99.96\\
20-Mult. & 7.30 & 24.11 & 39.40 & 59.24 & 66.34 & 72.63 & 76.71 & 80.26 & 83.23\\
\hline
\end{tabular} \centering
\label{tab:stackpercentages}
\end{table*}

\begin{table*}[ht]\centering
\footnotesize \centering \caption{The percentage of candidate GP
programs from the experiments conducted in this paper that can be
evaluated with a given stack size limit and an LGP representation
}
\begin{tabular}{|c|c|c|c|c|c|c|c|c|c|c|c|}
\hline
\multirow{2}{*}\textbf{Problem} & \multicolumn{11}{c|}{\textbf{Size of Stack Limit}}\\
\cline{2-12} & 1 & 2 & 3 & 4 & 5 & 6 & 7 & 8 & 9 & 10 & 11\\
\hline Sextic & 27.33 & 86.29 & 98.47 & 99.86 & 99.99 & 100.0 & 100.0 & 100.0 & 100.0 & 100.0 & 100.0\\
Shuttle & 25.61 & 87.06 & 97.77 & 99.57 & 99.91 & 99.98 & 100.0 & 100.0 & 100.0 & 100.0 & 100.0\\
KDDcup & 29.83 & 63.06 & 96.99 & 99.69 & 99.95 & 100.0 & 100.0 & 100.0 & 100.0 & 100.0 & 100.0\\
20-Mult. & 16.01 & 44.88 & 67.47 & 75.73 & 81.77 & 85.34 & 87.00 & 90.42 & 93.38 & 95.34 & 97.26\\
\hline
\end{tabular} \centering
\label{tab:LGPstack}
\end{table*}
\newpage
\section{Improving The Performance of a GPU Based 2D Stack For The
Purposes Of GP} \label{sec:2DGPGPUimprovements}In the previous
section it was observed that as the size of the second dimension
of the two dimensional stack increases, the number of threads of
execution which yielded the best computational performance
reduced. The key reason for this is that the level of on-chip L1
cache memory per stream processor is extremely limited. A modern
NVidia Kepler graphics card has 192 stream processors under each
SMX with only a maximum of 48KB of cache memory available. Shared
between the stream processors this is only 256 bytes per processor
or 64 32bit floating point values. Subsequently, to handle this
issue the pressure on the L1 cache must be reduced. Clearly, the
key methodology of achieving this would be to reduce the stack
operations that a candidate GP program requires. Indeed, a recent
study improved the performance of GPU based GP by reducing the
usage of the stack by modifying the program representation
\citep{Chitty:2014}.

With tree-based GP typically a postfix Reverse Polish Notation
(RPN) representation is used as this enables a faster iterative
interpreter to be used rather than a recursive interpreter. Thus,
using the RPN representation essentially a program is represented
in reverse with data values placed onto the stack and then
whenever a GP function is encountered, the required input values
are removed from the stack and the GP function executed upon these
inputs. However, with this approach, most of the stack operations
are essentially placing input values from the fitness cases onto
the stack only for them to be almost immediately removed again to
be utilised within a GP function. This unnecessarily increases the
number of stack operations and also unduly raises the amount of
stack memory required. Furthermore, this will also increase the
pressure upon the limited L1 cache.

An alternative approach would be to use a prefix GP representation
whereby whenever a GP function is interpreted, the inputs are
either directly accessed data values or constant values. In order
to facilitate this, a further input category is also required,
that of the result from a previously interpreted function which
will be on the stack. However, using a prefix representation for
candidate GP programs is complex if a hierarchical approach is
required such as that used in tree-based GP. If a GP function
requires an input from another function this input function would
need to be interpreted first and so forth. This would require a
fully recursive interpreter which calls itself rather than an
iterative interpreter which simply uses a loop and is thus much
faster. A better methodology of using a prefix representation to
candidate GP programs is to construct a list of functions to
execute. Thus a candidate GP program consists of a set of
functions whereby the inputs can take three values: a fitness case
data value, a constant value or a stack value. In the case of the
stack, this means the top of the stack whereby the result of the
previously interpreted GP function has been stored. The results
from the GP functions are always placed on the top of the stack.
Thus, using a prefix representation to candidate GP programs, if
the functions are interpreted in the right order the same result
as a postfix GP program representation can be achieved. This
representation is actually synonymous with Linear GP (LGP)
\cite{Brameier:2001}.

\begin{figure}
\centering
\epsfig{file=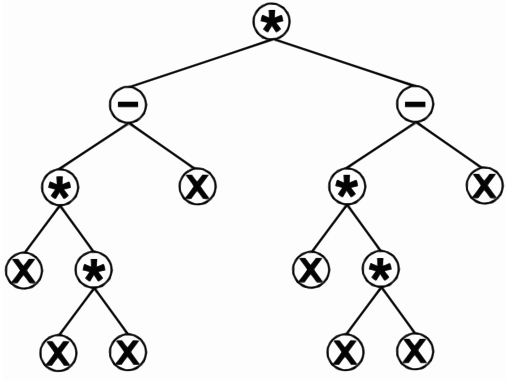,width=0.4\textwidth}
\caption{GP Tree representing the symbolic regression function
$x^6-2x^4+x^2$ which can be rewritten as $(x^3-x)*(x^3-x)$}
\label{fig:SymbolicRegressionTree}
\end{figure}

Consider Figure \ref{fig:SymbolicRegressionTree} which shows the
GP tree for the correct solution to the sextic regression problem.
This can be represented in postfix RPN as:
\\\\
\indent(((X,(X,X)+)*,X)-,((X,(X,X)+)*,X)-)*
\\\\
An alternative LGP form can be represented in prefix form as:
\\\\
\indent+(XX) *(XS) -(SX) +(XX) *(XS) -(SX) *(SS)
\\\\
where X is a fitness case input and S indicates a data input from
the top level of the stack. Outputs from functions are placed upon
the top of the stack. This LGP form of representation can be
considered as a simple list of instructions that can be
interpreted from left to right. From this it can be observed that
the LGP representation is larger in size with 21 values but there
are effectively only seven instructions for the interpreter to
execute versus fifteen instructions for the postfix RPN form. This
means that the LGP approach requires only seven cycles through the
instruction interpreter compared to fifteen cycles by the
traditional GP interpreter. Moreover, only six stack fetches are
performed compared to the RPN approach which has fourteen, two per
function. Furthermore, the maximum stack level required for the
LGP representation is only two compared to four for the postfix
RPN representation. Thus the efficiency savings from using an LGP
representation are potentially quite large using this approach.
Firstly, there are less stack operations thus less memory
operations which are slow. Secondly, there are less iterations
through the interpreter although the number of conditionals
evaluated are not dissimilar. Thirdly, the amount of stack memory
required is lower which reduces the level of L1 cache misses.

To demonstrate the savings in the required stack memory that can
be achieved, the percentage of candidate GP programs that can be
evaluated using a given stack limit for the experiments from
Section \ref{sec:2DGP} are shown for both representations. Table
\ref{tab:stackpercentages} shows the percentage of programs that
can be evaluated using a given stack limit for each problem
instance when using the postfix RPN representation. Table
\ref{tab:LGPstack} shows the percentage of programs that can be
evaluated using a given stack limit when using an LGP
representation. Clearly, a greater number of candidate GP programs
can be evaluated for a given stack limit when using the LGP
representation. Indeed, using only a stack limit size of three
enables over 95\% of candidate GP programs to be evaluated for
three of the problem instances. Compare this to a postfix RPN
representation whereby only a maximum of 36\% of candidate GP
programs can be evaluated using a stack limit size of three.

\begin{minipage}{0.45\textwidth}
\begin{lstlisting}[caption=Modified Two Dimensional Stack GP Interpreter With Reduced Stack Usage,
  label=alg:LinearGP,basicstyle=\footnotesize,numberstyle=\footnotesize]
#define STACKDIM 2
__global__ void CUDAInterpreter(int* progs,
        float* inputs, float* outputs, float* results,
        int maxprogsize) {
    extern __shared__ float prog[];
    for(int i=0;i<maxprogsize;i++)
        prog[i]=progs[blockIdx.x*maxprogsize+i];
    float stack[50][STACKDIM];
    float Sum=0.0;
    for(int i=threadIdx.x;i<NumCases;
        i+=NUM_THRDS*STACKDIM)
    {
        int ip=0; int sp=0;
        register float x1,y1,x2,y2;
        while(prog[ip]!=255) {
            if (prog[ip]==ADDITION) {
                ip++;
                if (prog[ip]==STACK_VALUE) {
                    sp--;
                    x1=stack[sp][0];x2=stack[sp][1];
                }
                else if (prog[ip]==INPUT_VALUE) {
                    int p=prog[ip]*NumCases+i;
                    x1=inputs[p]; p+=NUM_THRDS;
                    x2=inputs[p];
                }
                else if (prog[ip]==CONST_VALUE) {
                    x1=prog[ip];x2=prog[ip];
                }
                ip++;
                if (prog[ip]==STACK_VALUE) {
                    sp--;
                    y1=stack[sp][0]; y2=stack[sp][1];
                }
                else if (prog[ip]==INPUT_VALUE) {
                    int p=prog[ip]*NumCases+i;
                    y1=inputs[p]; p+=NUM_THRDS;
                    y2=inputs[p];
                }
                else if (prog[ip]==CONST_VALUE) {
                    y1=prog[ip]; y2=prog[ip];
                }
                ip++; stack[sp][0]=x1+y1;
                stack[sp][1]=x2+y2; sp++;
            }
        }
        for(int j=0;j<STACKDIM;j++)
            Sum+=GetError(stack[0][j],
                outputs[i+(j*NUM_THRDS)]);
    }
    results[blockIdx.x*NUM_THRDS+threadIdx.x]=Sum;
}
\end{lstlisting}
\end{minipage}

\begin{table*}[ht]\centering
\footnotesize \centering \caption{The GPop/s (measured in
billions) for each problem instance using the \emph{BlockGP}
single dimensional stack approach with an LGP representation with
differing preferences for the on-chip memory and a range of
threads within a block. The best performance for each problem
instance is shown in bold.}
\begin{tabular}{|c|c|c|c|c|}
\hline
\multirow{2}{*}\textbf{Problem} & \multirow{2}{*}\textbf{Num.} & \multicolumn{3}{c|}{\textbf{On-Chip Memory Preference}}\\
\cline{3-5} & \textbf{Threads} & Shared & Equal & L1\\
\hline
& 128 & 27.653 $\pm$ 0.219 & 28.303 $\pm$ 0.226 & 28.111 $\pm$ 0.173\\
& 256 & 28.232 $\pm$ 0.204 & \textbf{28.873} $\pm$ 0.209 & 28.752 $\pm$ 0.174\\
Sextic & 384 & 27.737 $\pm$ 0.225 & 28.253 $\pm$ 0.179 & 28.131 $\pm$ 0.173\\
& 512 & 27.795 $\pm$ 0.186 & 28.373 $\pm$ 0.175 & 28.344 $\pm$ 0.173\\
& 640 & 27.011 $\pm$ 0.169 & 27.439 $\pm$ 0.182 & 27.456 $\pm$ 0.151\\
& 768 & 24.719 $\pm$ 0.118 & 24.804 $\pm$ 0.095 & 24.850 $\pm$ 0.075\\
& 896 & 25.795 $\pm$ 0.144 & 26.083 $\pm$ 0.152 & 26.084 $\pm$ 0.112\\
& 1024 & 26.433 $\pm$ 0.170 & 26.888 $\pm$ 0.123 & 26.827 $\pm$ 0.126\\

\hline
& 128 & 29.704 $\pm$ 0.158 & 30.412 $\pm$ 0.187 & 30.385 $\pm$ 0.198\\
& 256 & 31.020 $\pm$ 0.141 & \textbf{31.780} $\pm$ 0.205 & 31.751 $\pm$ 0.153\\
Shuttle & 384 & 30.347 $\pm$ 0.152 & 31.023 $\pm$ 0.191 & 31.027 $\pm$ 0.200\\
& 512 & 30.553 $\pm$ 0.197 & 31.230 $\pm$ 0.175 & 31.236 $\pm$ 0.146\\
& 640 & 29.425 $\pm$ 0.139 & 30.026 $\pm$ 0.192 & 30.031 $\pm$ 0.205\\
& 768 & 26.340 $\pm$ 0.142 & 26.622 $\pm$ 0.092 & 26.621 $\pm$ 0.111\\
& 896 & 27.791 $\pm$ 0.157 & 28.231 $\pm$ 0.132 & 28.265 $\pm$ 0.139\\
& 1024 & 28.540 $\pm$ 0.151 & 29.043 $\pm$ 0.160 & 29.073 $\pm$ 0.126\\

\hline
& 128 & 27.129 $\pm$ 0.043 & 31.801 $\pm$ 0.214 & 32.019 $\pm$ 0.186\\
& 256 & 29.043 $\pm$ 0.058 & 33.767 $\pm$ 0.244 & 34.107 $\pm$ 0.283\\
KDDcup & 384 & 30.364 $\pm$ 0.130 & 33.292 $\pm$ 0.272 & 33.464 $\pm$ 0.273\\
& 512 & 29.938 $\pm$ 0.076 & 34.586 $\pm$ 0.306 & \textbf{34.759} $\pm$ 0.290\\
& 640 & 30.554 $\pm$ 0.115 & 33.507 $\pm$ 0.272 & 33.621 $\pm$ 0.231\\
& 768 & 27.990 $\pm$ 0.154 & 29.345 $\pm$ 0.161 & 29.435 $\pm$ 0.151\\
& 896 & 30.198 $\pm$ 0.217 & 32.248 $\pm$ 0.201 & 32.357 $\pm$ 0.224\\
& 1024 & 29.940 $\pm$ 0.077 & 34.298 $\pm$ 0.250 & 34.608 $\pm$ 0.286\\

\hline
& 128 & 860.442 $\pm$ 3.190 & 809.751 $\pm$ 2.612 & 564.267 $\pm$ 1.032\\
& 256 & 930.843 $\pm$ 3.415 & 970.622 $\pm$ 3.016 & 845.456 $\pm$ 2.131\\
20-Mult. & 384 & 926.819 $\pm$ 3.052 & 965.081 $\pm$ 4.038 & 955.701 $\pm$ 3.465\\
& 512 & 944.770 $\pm$ 3.458 & 984.065 $\pm$ 2.862 & \textbf{992.630} $\pm$ 4.271\\
& 640 & 915.959 $\pm$ 2.689 & 955.420 $\pm$ 3.619 & 963.909 $\pm$ 2.563\\
& 768 & 834.625 $\pm$ 2.410 & 863.949 $\pm$ 3.334 & 869.733 $\pm$ 2.135\\
& 896 & 882.457 $\pm$ 2.707 & 918.402 $\pm$ 3.256 & 925.366 $\pm$ 3.143\\
& 1024 & 912.472 $\pm$ 2.937 & 948.529 $\pm$ 3.637 & 954.921 $\pm$ 3.864\\
\hline
\end{tabular} \centering
\label{tab:RobilliardDynamicLinear}
\end{table*}

It could be considered that LGP should be used throughout the GP
run. However, LGP uses differing crossover and mutation operators
to those of tree-based GP. Subsequently, it would be much harder
to compare the performance between the two approaches as
completely different candidate GP programs would be evaluated.
Moreover, according to a recent GP survey \citep{White:2013},
tree-based GP is used by 82\% of GP practitioners as opposed to
only 25\% who use LGP thus converting candidate GP programs into
an LGP representation offers the best of both types of GP.
Converting from a tree based postfix RPN representation to an LGP
representation is a straightforward process. Note from the two
representations shown earlier for the solution to the sextic
regression problem that the set of functions are in the same
order. Thus a \emph{symbolic} stack can be used whereby one of
three values is placed, a data input (X1..XN), a constant value,
or S to represent the top of the stack. Reading the postfix RPN
form from left to right, if a data input value or constant is
encountered these are placed upon the \emph{symbolic} stack.
Whenever a GP function is encountered, this is written to the LGP
representation followed by the top \emph{symbolic} stack values
according to the arity of the given GP function. Finally, the
value S is placed on the \emph{symbolic} stack to indicate the
output of a GP function. Being able to convert a tree based GP
representation to an LGP form enables a direct comparison of
interpreter efficiency to be made between the two as the same GP
trees will be evaluated. Crossover and mutation are performed on
the tree-based GP representation but evaluation is performed using
the LGP representation.

An interpreter which operates using a prefix LGP representation to
GP programs for a two dimensional stack size of two is shown in
Listing \ref{alg:LinearGP}. The inputs to a given GP function such
as addition are placed in registers x and y. As in Listing
\ref{alg:LinearGP} the two dimensional stack size is of size two,
there are four registers labelled x1, x2, y1 and y2 as shown on
line 14. In cases of larger two dimensional stack sizes a greater
number of registers are required. It can also now be observed that
upon each iteration of the main interpreter loop, only a GP
function is expected. Once which GP function to execute is
ascertained, the required inputs are gathered and placed into the
registers. These inputs can consist of data values from the
fitness cases, constant values or the top of the stack denoted by
\emph{STACK\_VALUE}. The stack now only contains the results of
previous GP functions so \emph{STACK\_VALUE} means the output from
the last GP function executed is to be used as an input value.
Once the inputs are gathered the GP function is executed over the
two fitness cases and the results are placed on the top of the
stack as shown on lines 43-44. Of further note is that the loops
that iterate over the two dimensional stack have been removed.
Instead separate lines of code perform operations such as a
executing a GP function or gathering data inputs from the fitness
cases. This is due to the use of individually specified registers
rather than an indexable array. This can be most clearly seen on
lines 23-25 whereby data values from the fitness cases are placed
into the two registers x1 and x2.

\begin{table*}[ht]\centering
\footnotesize \centering \caption{The GPop/s (measured in
billions) for each problem instance using a two dimensional stack
and an LGP representation with a range of threads within a block
and range of second dimension sizes. The best performance for each
problem instance is shown in bold.}
\begin{tabular}{|c|c|c|c|c|c|c|}
\hline
\multirow{2}{*}\textbf{Problem} & \multirow{2}{*}\textbf{Num.} & \multicolumn{5}{c|}{\textbf{Size of Second Dimension Of Stack}}\\
\cline{3-7} & \textbf{Threads} & 2 & 3 & 4 & 5 & 6\\
\hline & 32 & 15.037 $\pm$ 0.035 & 19.018 $\pm$ 0.056 & 21.263 $\pm$ 0.085 & 21.448 $\pm$ 0.055 & 23.093 $\pm$ 0.066\\
& 64 & 27.302 $\pm$ 0.157 & 33.833 $\pm$ 0.146 & 37.114 $\pm$ 0.154 & 36.636 $\pm$ 0.207 & 38.295 $\pm$ 0.236\\
& 96 & 36.264 $\pm$ 0.151 & 41.794 $\pm$ 0.191 & 43.590 $\pm$ 0.236 & 42.250 $\pm$ 0.270 & 42.342 $\pm$ 0.245\\
& 128 & 42.481 $\pm$ 0.221 & 44.931 $\pm$ 0.285 & 45.700 $\pm$ 0.436 & 44.535 $\pm$ 0.297 & 44.058 $\pm$ 0.319\\
& 256 & 42.892 $\pm$ 0.220 & 45.193 $\pm$ 0.255 & \textbf{45.758} $\pm$ 0.286 & 44.524 $\pm$ 0.269 & 43.692 $\pm$ 0.297\\
Sextic & 384 & 41.215 $\pm$ 0.217 & 43.986 $\pm$ 0.270 & 44.621 $\pm$ 0.352 & 43.781 $\pm$ 0.224 & 42.314 $\pm$ 0.241\\
& 512 & 41.144 $\pm$ 0.199 & 43.073 $\pm$ 0.218 & 44.273 $\pm$ 0.255 & 42.257 $\pm$ 0.245 & 40.955 $\pm$ 0.268\\
& 640 & 39.072 $\pm$ 0.203 & 41.369 $\pm$ 0.275 & 42.434 $\pm$ 0.220 & 40.357 $\pm$ 0.243 & 37.586 $\pm$ 0.172\\
& 798 & 34.081 $\pm$ 0.149 & 37.743 $\pm$ 0.162 & 39.117 $\pm$ 0.209 & 37.595 $\pm$ 0.193 & 38.168 $\pm$ 0.157\\
& 896 & 36.294 $\pm$ 0.168 & 38.511 $\pm$ 0.158 & 39.748 $\pm$ 0.197 & 37.560 $\pm$ 0.197 & 29.847 $\pm$ 0.104\\
& 1024 & 37.402 $\pm$ 0.151 & 38.648 $\pm$ 0.186 & 39.328 $\pm$ 0.180 & 37.619 $\pm$ 0.160 & 30.354 $\pm$ 0.121\\

\hline & 32 & 16.095 $\pm$ 0.058 & 21.691 $\pm$ 0.048 & 25.817 $\pm$ 0.147 & 26.488 $\pm$ 0.130 & 29.320 $\pm$ 0.128\\
& 64 & 28.703 $\pm$ 0.131 & 38.204 $\pm$ 0.269 & 43.881 $\pm$ 0.400 & 43.585 $\pm$ 0.374 & 46.049 $\pm$ 0.346\\
& 96 & 37.196 $\pm$ 0.254 & 47.909 $\pm$ 0.293 & 49.103 $\pm$ 0.225 & 45.060 $\pm$ 0.213 & 40.126 $\pm$ 0.126\\
& 128 & 43.339 $\pm$ 0.416 & 47.023 $\pm$ 0.165 & 38.133 $\pm$ 0.073 & 43.145 $\pm$ 0.114 & 38.440 $\pm$ 0.038\\
& 256 & 44.279 $\pm$ 0.337 & 48.203 $\pm$ 0.167 & 38.855 $\pm$ 0.053 & 43.549 $\pm$ 0.134 & 39.206 $\pm$ 0.060\\
Shuttle & 384 & 42.676 $\pm$ 0.313 & \textbf{49.734} $\pm$ 0.336 & 41.730 $\pm$ 0.075 & 43.427 $\pm$ 0.147 & 39.161 $\pm$ 0.059\\
& 512 & 42.661 $\pm$ 0.267 & 47.259 $\pm$ 0.180 & 38.853 $\pm$ 0.053 & 42.878 $\pm$ 0.131 & 39.814 $\pm$ 0.077\\
& 640 & 40.474 $\pm$ 0.343 & 46.705 $\pm$ 0.372 & 39.575 $\pm$ 0.068 & 43.345 $\pm$ 0.351 & 42.009 $\pm$ 0.159\\
& 798 & 36.215 $\pm$ 0.209 & 43.675 $\pm$ 0.324 & 44.724 $\pm$ 0.204 & 40.412 $\pm$ 0.160 & 37.144 $\pm$ 0.099\\
& 896 & 37.486 $\pm$ 0.277 & 44.099 $\pm$ 0.238 & 39.807 $\pm$ 0.100 & 36.296 $\pm$ 0.254 & 37.182 $\pm$ 0.178\\
& 1024 & 38.269 $\pm$ 0.309 & 42.062 $\pm$ 0.188 & 35.872 $\pm$ 0.041 & 36.290 $\pm$ 0.281 & 37.032 $\pm$ 0.162\\

\hline & 32 & 14.779 $\pm$ 0.005 & 20.400 $\pm$ 0.005 & 24.364 $\pm$ 0.007 & 25.115 $\pm$ 0.012 & 27.098 $\pm$ 0.013 \\
& 64 & 27.387 $\pm$ 0.009 & 37.024 $\pm$ 0.012 & 38.646 $\pm$ 0.045 & 32.659 $\pm$ 0.033 & 27.738 $\pm$ 0.035 \\
& 96 & 36.462 $\pm$ 0.021 & 33.800 $\pm$ 0.027 & 29.894 $\pm$ 0.025 & 30.354 $\pm$ 0.014 & 26.368 $\pm$ 0.018\\
& 128 & 33.898 $\pm$ 0.063 & 33.284 $\pm$ 0.032 & 29.341 $\pm$ 0.020 & 29.785 $\pm$ 0.012 & 25.822 $\pm$ 0.007\\
& 256 & 35.422 $\pm$ 0.025 & 34.406 $\pm$ 0.022 & 30.034 $\pm$ 0.016 & 30.797 $\pm$ 0.007 & 26.307 $\pm$ 0.023\\
KDDcup & 384 & 38.540 $\pm$ 0.018 & 35.488 $\pm$ 0.012 & 30.971 $\pm$ 0.008 & 32.974 $\pm$ 0.014 & 28.117 $\pm$ 0.016\\
& 512 & 37.269 $\pm$ 0.021 & 36.147 $\pm$ 0.012 & 31.162 $\pm$ 0.011 & 35.854 $\pm$ 0.020 & 29.626 $\pm$ 0.016\\
& 640 & 38.574 $\pm$ 0.029 & 40.903 $\pm$ 0.025 & 34.408 $\pm$ 0.016 & 30.925 $\pm$ 0.009 & 26.049 $\pm$ 0.005\\
& 798 & 40.091 $\pm$ 0.015 & 35.562 $\pm$ 0.035 & 30.972 $\pm$ 0.020 & 36.895 $\pm$ 0.006 & 33.321 $\pm$ 0.024\\
& 896 & \textbf{41.102} $\pm$ 0.074 & 36.674 $\pm$ 0.007 & 40.543 $\pm$ 0.060 & 38.490 $\pm$ 0.087 & 30.183 $\pm$ 0.010\\
& 1024 & 37.176 $\pm$ 0.033 & 39.970 $\pm$ 0.022 & 40.708 $\pm$ 0.023 & 35.423 $\pm$ 0.008 & 28.822 $\pm$ 0.009\\

\hline & 32 & 257.2 $\pm$ 0.412 & 328.126 $\pm$ 0.305 & 400.714 $\pm$ 0.612 & 399.403 $\pm$ 0.742 & 452.839 $\pm$ 0.922\\
& 64 & 483.842 $\pm$ 0.859 & 610.438 $\pm$ 1.436 & 728.638 $\pm$ 1.788 & 717.745 $\pm$ 2.052 & 794.417 $\pm$ 2.114\\
& 96 & 674.451 $\pm$ 2.068 & 835.056 $\pm$ 2.919 & 955.598 $\pm$ 3.165 & 928.226 $\pm$ 2.095 & 989.579 $\pm$ 1.897\\
& 128 & 844.012 $\pm$ 2.660 & 996.948 $\pm$ 3.461 & 1081.180 $\pm$ 1.956 & 1020.482 $\pm$ 1.590 & 1052.673 $\pm$ 1.135\\
& 256 & 1182.924 $\pm$ 5.157 & 1220.506 $\pm$ 5.019 & 1140.245 $\pm$ 2.159 & 1039.301 $\pm$ 1.607 & 980.096 $\pm$ 1.268\\
20-Mult. & 384 & \textbf{1269.799} $\pm$ 4.924 & 1134.336 $\pm$ 4.274 & 1016.856 $\pm$ 2.497 & 687.620 $\pm$ 0.401 & 898.617 $\pm$ 1.365\\
& 512 & 1258.405 $\pm$ 5.703 & 1135.297 $\pm$ 4.816 & 1005.648 $\pm$ 3.175 & 933.954 $\pm$ 1.749 & 889.863 $\pm$ 1.504\\
& 640 & 1237.728 $\pm$ 4.617 & 1094.886 $\pm$ 4.698 & 987.487 $\pm$ 1.634 & 894.353 $\pm$ 2.466 & 873.260 $\pm$ 1.868\\
& 798 & 1153.680 $\pm$ 5.530 & 1155.208 $\pm$ 4.923 & 1074.670 $\pm$ 3.915 & 981.970 $\pm$ 3.022 & 931.425 $\pm$ 2.713\\
& 896 & 1194.155 $\pm$ 5.928 & 1090.577 $\pm$ 5.043 & 990.088 $\pm$ 2.585 & 896.312 $\pm$ 2.282 & 877.632 $\pm$ 2.381\\
& 1024 & 1186.163 $\pm$ 3.233 & 1056.634 $\pm$ 3.668 & 966.358 $\pm$ 2.234 & 871.310 $\pm$ 2.125 & 828.399 $\pm$ 1.424\\
\hline
\end{tabular} \centering
\label{tab:2DstackGPULinear}
\end{table*}
\newpage
In order to test the performance advantage obtained by using an
LGP conversion of the tree-based GP candidate GP programs, the
experiments that generated the results in Tables
\ref{tab:RobilliardDynamic} and \ref{tab:2DstackGPU} will be
repeated now using the LGP representation. These results are shown
in Tables \ref{tab:RobilliardDynamicLinear} and
\ref{tab:2DstackGPULinear}. Comparing the single dimensional stack
\emph{BlockGP} RPN and LGP representations in Tables
\ref{tab:RobilliardDynamic} and \ref{tab:RobilliardDynamicLinear}
it can be observed that a significant performance gain has been
achieved with much faster results from using an LGP
representation. For the first three problem instances an average
of 31.8 billion GPop/s have now been achieved with a peak of 34.7
billion GPop/s for the KDDcup classification problem. For the
multiplexer problem instance nearly one trillion GPop/s have been
achieved. An average performance gain of 1.17x over the
traditional tree-based GP approach has been achieved using an LGP
representation.

However, the results for using a two dimensional stack approach
with an LGP representation are more impressive. For the first
three problem instances an average performance of 45.5 billion
GPop/s are achieved with a peak of nearly 50 billion GPop/s for
the Shuttle classification problem. With the multiplexer problem,
1.27 trillion GPop/s have been achieved. The average performance
gain of the two dimensional stack LGP representation approach over
the RPN representation is 1.44x. A further observation can be made
from Table \ref{tab:2DstackGPULinear} in that the number of
fitness cases that can be considered simultaneously has increased
for two of the problem instances to three fitness cases. However,
it can still be observed that as the size of the second dimension
of the stack increases, the best computational performance is
achieved with fewer threads although the effect is much more
reduced when using the LGP representation. Comparing the two
dimensional stack approach to a single dimensional stack when
using an LGP representation, an average performance gain of 1.4x
is achieved. When comparing the LGP two dimensional GP approach to
a single dimensional stack with an RPN representation, an average
performance gain of 1.64x has now been achieved.

\begin{table*}[ht]\centering
\footnotesize \centering \caption{The GPop/s (measured in
billions) for each problem instance using an LGP conversion and
differing extended data types representing differing dimension
sizes for the two dimensional stack approach with a range of
threads within a block. The best performance for each problem
instance is shown in bold.}
\begin{tabular}{|c|c|c|c|c|}
\hline
\multirow{2}{*}\textbf{Problem} & \multirow{2}{*}\textbf{Num.} & \multicolumn{3}{c|}{\textbf{Data Type}}\\
\cline{3-5} & \textbf{Threads} & Float2 & Float3 & Float4\\
\hline & 32 & 15.784 $\pm$ 0.053 & 19.699 $\pm$ 0.063 & 22.493 $\pm$ 0.037\\
& 64 & 29.019 $\pm$ 0.116 & 34.726 $\pm$ 0.178 & 38.529 $\pm$ 0.349\\
& 96 & 38.163 $\pm$ 0.226 & 43.021 $\pm$ 0.234 & 44.818 $\pm$ 0.227\\
& 128 & 44.470 $\pm$ 0.429 & 46.767 $\pm$ 0.232 & 46.746 $\pm$ 0.308\\
& 256 & 45.483 $\pm$ 0.309 & \textbf{47.065} $\pm$ 0.315 & 46.952 $\pm$ 0.273\\
Sextic & 384 & 43.301 $\pm$ 0.261 & 45.929 $\pm$ 0.249 & 46.096 $\pm$ 0.298\\
& 512 & 43.504 $\pm$ 0.274 & 45.126 $\pm$ 0.241 & 45.572 $\pm$ 0.260\\
& 640 & 41.655 $\pm$ 0.242 & 43.396 $\pm$ 0.289 & 44.047 $\pm$ 0.248\\
& 768 & 37.152 $\pm$ 0.240 & 39.740 $\pm$ 0.217 & 41.372 $\pm$ 0.209\\
& 896 & 39.121 $\pm$ 0.237 & 40.369 $\pm$ 0.214 & 41.752 $\pm$ 0.202\\
& 1024 & 39.971 $\pm$ 0.216 & 40.407 $\pm$ 0.216 & 41.317 $\pm$ 0.240\\

\hline & 32 & 16.865 $\pm$ 0.056 & 21.701 $\pm$ 0.072 & 25.892 $\pm$ 0.138\\
& 64 & 30.567 $\pm$ 0.159 & 37.974 $\pm$ 0.258 & 43.234 $\pm$ 0.379\\
& 96 & 40.116 $\pm$ 0.285 & 47.844 $\pm$ 0.476 & 48.885 $\pm$ 0.278\\
& 128 & 46.904 $\pm$ 0.493 & 46.249 $\pm$ 0.154 & 38.829 $\pm$ 0.072\\
& 256 & 48.009 $\pm$ 0.351 & 47.482 $\pm$ 0.210 & 39.481 $\pm$ 0.080\\
Shuttle & 384 & 46.108 $\pm$ 0.539 & \textbf{49.528} $\pm$ 0.242 & 42.731 $\pm$ 0.118\\
& 512 & 46.051 $\pm$ 0.445 & 47.167 $\pm$ 0.161 & 39.357 $\pm$ 0.099\\
& 640 & 43.642 $\pm$ 0.318 & 46.567 $\pm$ 0.333 & 40.175 $\pm$ 0.099\\
& 768 & 38.732 $\pm$ 0.283 & 43.788 $\pm$ 0.328 & 44.464 $\pm$ 0.214\\
& 896 & 40.218 $\pm$ 0.282 & 44.079 $\pm$ 0.292 & 40.072 $\pm$ 0.236\\
& 1024 & 40.953 $\pm$ 0.356 & 41.902 $\pm$ 0.203 & 36.364 $\pm$ 0.056\\

\hline & 32 & 14.847 $\pm$ 0.002 & 18.993 $\pm$ 0.007 & 23.460 $\pm$ 0.007\\
& 64 & 28.126 $\pm$ 0.011 & 34.597 $\pm$ 0.032 & 38.832 $\pm$ 0.029\\
& 96 & 37.342 $\pm$ 0.088 & 32.843 $\pm$ 0.021 & 30.364 $\pm$ 0.027\\
& 128 & 33.610 $\pm$ 0.017 & 27.901 $\pm$ 0.016 & 29.815 $\pm$ 0.015\\
& 256 & 35.260 $\pm$ 0.062 & 28.756 $\pm$ 0.038 & 30.540 $\pm$ 0.010\\
KDDcup & 384 & 38.467 $\pm$ 0.062 & 30.782 $\pm$ 0.043 & 31.613 $\pm$ 0.010\\
& 512 & 37.182 $\pm$ 0.021 & 30.351 $\pm$ 0.046 & 31.779 $\pm$ 0.010\\
& 640 & 38.700 $\pm$ 0.024 & 30.617 $\pm$ 0.007 & 34.858 $\pm$ 0.018\\
& 768 & 41.376 $\pm$ 0.125 & 36.076 $\pm$ 0.019 & 31.659 $\pm$ 0.012\\
& 896 & \textbf{41.555} $\pm$ 0.016 & 31.906 $\pm$ 0.031 & 39.234 $\pm$ 0.020\\
& 1024 & 37.155 $\pm$ 0.037 & 29.969 $\pm$ 0.009 & 40.700 $\pm$ 0.028\\

\hline & 32 & 256.427 $\pm$ 0.197 & 332.988 $\pm$ 0.415 & 398.895 $\pm$ 0.779\\
& 64 & 486.622 $\pm$ 0.916 & 619.829 $\pm$ 0.993 & 719.988 $\pm$ 2.062\\
& 96 & 683.578 $\pm$ 1.911 & 848.797 $\pm$ 2.697 & 936.386 $\pm$ 3.240\\
& 128 & 852.527 $\pm$ 2.486 & 1008.896 $\pm$ 2.871 & 1062.896 $\pm$ 2.431\\
& 256 & 1189.307 $\pm$ 5.929 & 1219.041 $\pm$ 5.221 & 1131.851 $\pm$ 2.959\\
20-Mult. & 384 & \textbf{1263.537} $\pm$ 4.459 & 1128.878 $\pm$ 3.734 & 1022.320 $\pm$ 3.168\\
& 512 & 1242.809 $\pm$ 6.919 & 1107.963 $\pm$ 5.070 & 1014.055 $\pm$ 2.691\\
& 640 & 1226.406 $\pm$ 5.213 & 1087.256 $\pm$ 4.405 & 989.376 $\pm$ 2.333\\
& 768 & 1155.696 $\pm$ 5.002 & 1156.151 $\pm$ 4.025 & 1075.118 $\pm$ 3.091\\
& 896 & 1183.008 $\pm$ 4.813 & 1088.237 $\pm$ 4.821 & 993.639 $\pm$ 2.808\\
& 1024 & 1169.623 $\pm$ 4.048 & 1044.892 $\pm$ 4.051 & 966.918 $\pm$ 2.750\\
\hline
\end{tabular} \centering
\label{tab:2DstackGPULinearSSE}
\end{table*}

\section{Further Improvements To The GPU 2D Stack Model} \label{sec:2DGPGPUFurtherImprovements}
\subsection{Utilising SSE Data Types} \label{sec:2DGPGPUsse}The previous section has
demonstrated that a two dimensional stack model provides a
significant computational performance advantage over a single
dimensional stack for GPU based GP. Subsequently, each GP
operation is now performed over multiple fitness cases. This
approach can now enable a further form of data parallelism to be
used. Indeed, in the original work which implemented a two
dimensional stack for GP using a CPU \citep{Chitty:2012}, the
performance was extended further by taking advantage of Streaming
SIMD Extensions (SSE) instructions. On a modern CPU there are a
number of larger registers up to 256 bits wide. This enables
multiple 32 bit floats to be stored in the same registers.
Subsequently, whenever an operation is performed on these wider
registers, the operation is performed upon several 32 bit floats
simultaneously. This can be considered a form of data parallelism.

The GPU programming language CUDA also has a set of data types
which can store more than one floating point value. These are
known as float2, float3 and float4 which can hold 2, 3 and 4 32
bit floats respectively. However, unlike a modern CPU, there are
not any extra wide registers on GPUs to enable simultaneous
parallel operations to be performed on these data types.
Operations must still be performed in a sequential manner.
However, using these registers can effectively simplify a CUDA
program and enable the source code to be more predictable to the
CUDA compiler which could provide a performance advantage as
instructions can then be better preloaded.

A float2 data type has two 32 bit float members x and y. Float3
has an additional float member z and a float4 data type has a
further member w. An addition operation on two float2 data types
involves adding the two x members together and then the two y
members. A simple inline operator function can be used for GP
functions such as addition an example of which is shown in Listing
\ref{alg:InlineFloat4}. Additionally, when accessing the stack, as
both the registers and the local memory stack are defined as
float4 datatypes, a simple assignment statement works in the same
manner as with a normal float datatype.

\begin{minipage}{0.45\textwidth}
\begin{lstlisting}[caption=Code segment demonstrating an inline addition GP operator using a float4 datatype,
  label=alg:InlineFloat4,basicstyle=\footnotesize,numberstyle=\footnotesize]
inline __host__ __device__
    float4 operator+(float4 a, float4 b)
{
    return make_float4(a.x+b.x, a.y+b.y,
        a.z+b.z, a.w+b.w)
}
\end{lstlisting}
\end{minipage}

The results from using these extended data types are shown in
Table \ref{tab:2DstackGPULinearSSE}. However, performance is only
slightly improved over the results from Table
\ref{tab:2DstackGPULinear}. Indeed, a performance advantage is
only observed for two of the problem instances whereby a float3
data type is most effective. Overall the average performance gain
was only 1.01x over the results from Table
\ref{tab:2DstackGPULinear}. However, this is as expected as
although there are extended data types, there are no underlying
extended registers. Therefore there is no data parallelism
performed although the source code of the interpreter is more
predictable and this aids the CUDA compiler which has led to the
small performance advantage that is observed. However, the use of
extended data types will be of benefit for the work presented in
the next section.

\subsection{Using The Register File For The Stack} \label{sec:2DGPGPUsseRegisters}Even though a
considerable speedup has been achieved by the work presented in
the previous sections, it is clear that the L1 cache memory is
still under pressure restricting the effectiveness of a two
dimensional stack model for GPU based GP. Subsequently, further
relieving the pressure on the L1 cache memory would be favorable
as it may enable more fitness cases to be considered in a single
pass of the interpreter. To achieve this a different memory
resource must be considered which is not cached but operates as
fast as L1 cache memory. In fact, the fastest memory resource
available on a GPU is the set of registers that are also located
on-chip. Indeed, each SMX has 64k of addressable registers
available to the threads executing under it which is a
considerable memory storage area but is often overlooked when
implementing algorithms for GPUs. In fact, Chitty
\citeyearpar{Chitty:2014} demonstrated that a considerable
improvement in the speed of single dimensional stack GP could be
achieved when utilising the GPU registers as part of the stack.

A key drawback in using registers to hold the stack is that they
are not indexable. Instead, a conditional \texttt{switch}
statement is required to ensure that the correct register is
accessed when performing a stack operation. This will inevitably
add an overhead to the operation of GP. However, since GP can be
considered as much \emph{memory bound} as \emph{compute bound} due
to the high level of stack operations and the limited memory
resources of GPUs, the benefit of improved memory operations
should outweigh this extra computational cost. An additional issue
is that by increasing the number of registers used by a single
thread of execution, the maximum number of threads that can be
executed in parallel by the GPU could be reduced impacting on
overall performance.

\begin{table*}[ht]\centering
\footnotesize \centering \caption{The GPop/s (measured in
billions) for each problem instance with the two dimensional stack
model using a float4 data type. A range of threads within a block
and a differing range of registers as part of the stack are shown.
The best performance for each problem instance is shown in bold.}
\begin{tabular} {|c|c|c|c|c|c|}
\hline
\multirow{2}{*}\textbf{Problem} & \multirow{2}{*}\textbf{Num.} & \multicolumn{4}{c|}{\textbf{Number Of Registers}}\\
\cline{3-6} & \textbf{Threads} & 1 & 2 & 3 & 4\\
\hline & 32 & 21.681 $\pm$ 0.051 & 22.350 $\pm$ 0.069 & 21.177 $\pm$ 0.098 & 21.025 $\pm$ 0.055\\
& 64 & 38.410 $\pm$ 0.266 & 39.588 $\pm$ 0.218 & 38.088 $\pm$ 0.246 & 37.531 $\pm$ 0.211\\
& 96 & 47.655 $\pm$ 0.267 & 48.310 $\pm$ 0.480 & 47.060 $\pm$ 0.340 & 41.912 $\pm$ 0.176\\
& 128 & 53.068 $\pm$ 0.323 & 51.223 $\pm$ 0.339 & 50.059 $\pm$ 0.296 & 44.463 $\pm$ 0.392\\
& 256 & \textbf{53.193} $\pm$ 0.388 & 51.224 $\pm$ 0.317 & 50.264 $\pm$ 0.209 & 44.439 $\pm$ 0.231\\
Sextic & 384 & 51.342 $\pm$ 0.434 & 49.767 $\pm$ 0.357 & 48.945 $\pm$ 0.334 & 41.145 $\pm$ 0.296\\
& 512 & 50.885 $\pm$ 0.298 & 48.498 $\pm$ 0.336 & 47.994 $\pm$ 0.365 & 37.931 $\pm$ 0.184\\
& 640 & 48.496 $\pm$ 0.351 & 43.149 $\pm$ 0.217 & 42.194 $\pm$ 0.264 & 41.222 $\pm$ 0.249\\
& 768 & 44.099 $\pm$ 0.192 & 45.506 $\pm$ 0.245 & 44.860 $\pm$ 0.368 & 29.591 $\pm$ 0.097\\
& 896 & 45.399 $\pm$ 0.263 & 34.248 $\pm$ 0.139 & 33.087 $\pm$ 0.203 & 32.588 $\pm$ 0.237\\
& 1024 & 45.364 $\pm$ 0.250 & 36.099 $\pm$ 0.189 & 35.089 $\pm$ 0.103 & 34.501 $\pm$ 0.135\\

\hline & 32 & 23.800 $\pm$ 0.121 & 23.938 $\pm$ 0.077 & 23.454 $\pm$ 0.114 & 23.138 $\pm$ 0.096\\
& 64 & 41.735 $\pm$ 0.283 & 42.183 $\pm$ 0.371 & 41.372 $\pm$ 0.361 & 40.540 $\pm$ 0.271\\
& 96 & 51.433 $\pm$ 0.476 & 52.150 $\pm$ 0.626 & 46.534 $\pm$ 0.467 & 45.479 $\pm$ 0.423\\
& 128 & 54.781 $\pm$ 0.540 & 55.555 $\pm$ 0.713 & 49.809 $\pm$ 0.470 & 48.717 $\pm$ 0.410\\
& 256 & 54.915 $\pm$ 0.579 & \textbf{55.734} $\pm$ 0.496 & 49.994 $\pm$ 0.300 & 48.744 $\pm$ 0.463\\
Shuttle & 384 & 53.563 $\pm$ 0.491 & 54.268 $\pm$ 0.420 & 45.863 $\pm$ 0.346 & 44.946 $\pm$ 0.342\\
& 512 & 51.398 $\pm$ 0.390 & 52.089 $\pm$ 0.373 & 41.374 $\pm$ 0.286 & 40.556 $\pm$ 0.347\\
& 640 & 45.757 $\pm$ 0.338 & 46.078 $\pm$ 0.464 & 45.475 $\pm$ 0.302 & 44.357 $\pm$ 0.367\\
& 768 & 48.134 $\pm$ 0.425 & 48.787 $\pm$ 0.598 & 32.577 $\pm$ 0.155 & 32.128 $\pm$ 0.202\\
& 896 & 34.742 $\pm$ 0.236 & 35.041 $\pm$ 0.240 & 34.565 $\pm$ 0.229 & 33.991 $\pm$ 0.230\\
& 1024 & 36.641 $\pm$ 0.198 & 36.992 $\pm$ 0.289 & 36.530 $\pm$ 0.218 & 35.861 $\pm$ 0.181\\

\hline & 32 & 22.119 $\pm$ 0.006 & 22.101 $\pm$ 0.004 & 21.735 $\pm$ 0.006 & 21.343 $\pm$ 0.004\\
& 64 & 40.069 $\pm$ 0.025 & 40.738 $\pm$ 0.016 & 40.245 $\pm$ 0.015 & 39.199 $\pm$ 0.072\\
& 96 & 37.154 $\pm$ 0.025 & 44.624 $\pm$ 0.045 & 44.465 $\pm$ 0.030 & 43.221 $\pm$ 0.025\\
& 128 & 36.925 $\pm$ 0.056 & 43.418 $\pm$ 0.025 & 47.963 $\pm$ 0.059 & 46.566 $\pm$ 0.110\\
& 256 & 37.911 $\pm$ 0.041 & 44.530 $\pm$ 0.048 & \textbf{48.811} $\pm$ 0.087 & 47.539 $\pm$ 0.084\\
KDDcup & 384 & 38.691 $\pm$ 0.021 & 45.473 $\pm$ 0.026 & 46.446 $\pm$ 0.028 & 45.276 $\pm$ 0.038\\
& 512 & 39.344 $\pm$ 0.019 & 46.274 $\pm$ 0.028 & 43.566 $\pm$ 0.109 & 42.698 $\pm$ 0.021\\
& 640 & 42.917 $\pm$ 0.014 & 48.750 $\pm$ 0.047 & 48.738 $\pm$ 0.132 & 47.618 $\pm$ 0.029\\
& 768 & 38.943 $\pm$ 0.015 & 45.946 $\pm$ 0.124 & 35.219 $\pm$ 0.017 & 34.637 $\pm$ 0.007\\
& 896 & 39.765 $\pm$ 0.008 & 40.167 $\pm$ 0.020 & 39.734 $\pm$ 0.073 & 39.067 $\pm$ 0.014\\
& 1024 & 43.259 $\pm$ 0.020 & 43.998 $\pm$ 0.028 & 43.638 $\pm$ 0.039 & 42.868 $\pm$ 0.043\\

\hline & 32 & 334.865 $\pm$ 0.411 & 352.504 $\pm$ 0.392 & 337.995 $\pm$ 0.443 & 329.912 $\pm$ 0.341\\
& 64 & 619.352 $\pm$ 1.665 & 652.192 $\pm$ 1.422 & 629.691 $\pm$ 1.278 & 616.685 $\pm$ 1.346\\
& 96 & 836.978 $\pm$ 2.021 & 885.453 $\pm$ 2.445 & 859.410 $\pm$ 2.985 & 832.178 $\pm$ 3.212\\
& 128 & 1016.015 $\pm$ 4.115 & 1079.990 $\pm$ 4.567 & 1047.476 $\pm$ 3.139 & 1016.706 $\pm$ 4.420\\
& 256 & 1254.759 $\pm$ 4.384 & 1378.092 $\pm$ 6.138 & 1386.533 $\pm$ 6.740 & 1330.446 $\pm$ 6.889\\
20-Mult. & 384 & 1173.130 $\pm$ 6.597 & 1326.854 $\pm$ 7.343 & \textbf{1411.185} $\pm$ 7.914 & 1257.478 $\pm$ 5.566\\
& 512 & 1138.594 $\pm$ 4.537 & 1300.657 $\pm$ 5.287 & 1401.609 $\pm$ 7.277 & 1173.107 $\pm$ 4.979\\
& 640 & 1127.237 $\pm$ 5.910 & 1278.740 $\pm$ 4.737 & 1273.437 $\pm$ 4.796 & 1272.889 $\pm$ 6.368\\
& 768 & 1169.368 $\pm$ 3.726 & 1275.609 $\pm$ 4.481 & 1318.829 $\pm$ 5.718 & 917.335 $\pm$ 3.534\\
& 896 & 1112.279 $\pm$ 5.120 & 1236.303 $\pm$ 5.199 & 976.641 $\pm$ 3.171 & 974.130 $\pm$ 3.622\\
& 1024 & 1083.003 $\pm$ 4.551 & 1215.954 $\pm$ 6.038 & 1068.662 $\pm$ 4.921 & 1070.943 $\pm$ 4.691\\
\hline
\end{tabular} \centering
\label{tab:2DstackGPULinearSSEregister}
\end{table*}

Since there is the potential that a given candidate GP program
could need a large stack, using registers to hold the whole stack
is considered infeasible. However, a mix of registers and local
memory could be used whereby the first $n$ levels of the stack use
registers and subsequent levels use the local memory stack which
benefits from the fast on-chip L1 cache memory. It could be
considered from Table \ref{tab:LGPstack} that the first few levels
of stack memory are the most important as most candidate GP
programs can be evaluated using a small stack. Subsequently, these
lowest levels of the stack should be represented by registers.

\begin{minipage}{0.45\textwidth}
\begin{lstlisting}[caption=Code segment demonstrating register file stack usage and getting a value from the stack,
  label=alg:RegisterStackGet,basicstyle=\footnotesize,numberstyle=\footnotesize]
register float4 x,y;
register float4 StackReg1,StackReg2;
float4 stack[50];
if (prog[ip]==STACK_VALUE) {
    sp--;
    switch(sp) {
        case 0: x=StackReg1;break;
        case 1: x=StackReg2;break;
        default: x=stack[sp];
    }
}
\end{lstlisting}
\end{minipage}

\begin{minipage}{0.45\textwidth}
\begin{lstlisting}[caption=Code segment demonstrating register file usage and placing a result on the stack after an addition operation,
  label=alg:RegisterStackPut,basicstyle=\footnotesize,numberstyle=\footnotesize]
register float4 x,y;
register float4 StackReg1,StackReg2;
float4 stack[50];

x=x+y;
switch(sp) {
    case 0: StackReg1=x;break;
    case 1: StackReg2=x;break;
    default: stack[sp]=x;
}
sp++;
\end{lstlisting}
\end{minipage}

Code segments that represent stack operations using two registers
to hold the first two levels of the stack are shown in Listings
\ref{alg:RegisterStackGet} and \ref{alg:RegisterStackPut}. A two
dimensional stack is used through the use of the float4 data type
enabling multiple fitness cases to be considered in a single
interpretation of a GP candidate solution. Listing
\ref{alg:RegisterStackGet} shows the switch statement required
when obtaining an input value from the stack. Listing
\ref{alg:RegisterStackPut} demonstrates how the result of a GP
addition operator is placed on the stack using a switch statement.

The results from using differing numbers of registers to hold the
first few levels of the stack are shown in Table
\ref{tab:2DstackGPULinearSSEregister}. In this case, only the
results from using a second dimension stack size of four are shown
which enables the use of the float4 extended data type.
Experiments were conducted with smaller two dimensional stack
sizes but the results were less effective than those shown in
Table \ref{tab:2DstackGPULinearSSEregister}. Across the four
problem instances the results are mixed, for the sextic regression
problem the best performance comes from using a single register,
the Shuttle classification problem two registers and the remaining
two problem instances, three registers. A possible reason for this
could be the usage of particular levels of the stack. If for a
given problem case the candidate GP solutions evaluated mostly
only use the first level of the stack then using extra registers
could have a detrimental effect on performance. This correlates
with Table \ref{tab:LGPstack} whereby for the sextic regression
and Shuttle classification problem instances, a majority of
candidate GP programs can be evaluated using only two stack
levels. Using greater numbers of registers can reduce the level of
parallelism that can be achieved as there are a limited number.
Moreover, the use of a larger number of registers to represent the
stack increases the size of the conditional \texttt{switch}
statement. However, from the results it could be considered that
the best number of registers to use is two with 256 threads of
execution per block. Using the first two levels of the stack
represented by registers and 256 threads of execution the average
performance gain over the results in Table
\ref{tab:2DstackGPULinearSSE} is 1.09x. Using the peak rates
achieved for each problem instance a performance gain of 1.14x is
achieved. Thus the use of registers to represent part of the stack
has provided a performance gain. Moreover, through the use of
registers, the size of the second dimension of the stack can be
extended to enable four fitness cases to be considered at each
step of a candidate program. This is possible as the use of
registers has further reduced the pressure on the L1 cache memory.
Overall, an average performance gain of 1.88x has been achieved
over the standard single dimensional stack GPU based
\emph{BlockGP} approach with a peak 2.11x performance gain. A
maximum rate of 55.7 billion GPop/s has been achieved for the
Shuttle classification problem and 1.4 trillion GPop/s for the
multiplexer problem which benefits from an extra level of bitwise
parallelism.

\section{Conclusions} \label{sec:Conclusions}
This paper has taken a two dimensional stack model to GP that was
applied with great success to a CPU architecture and applied it to
a GPU implementation of GP. By using a two dimensional stack
multiple fitness cases can be considered at each step of a program
thereby reducing interpreter overheads and thus improving the
execution speed. The dimensional stack model for GPU based GP is
more limited than the CPU implementation due to both the massive
level of parallelism available on GPUs vs CPUs and also the
reduced amount of L1 cache memory available on GPUs. However, even
given this limited model, improvements in the computational speed
were achieved by the two dimensional stack GP model over the best
single dimensional GP model from the literature, \emph{BlockGP}.

However, to achieve the best performance from the two dimensional
stack model stack operations need to be reduced to relieve the
pressure on the L1 cache. This was achieved by converting GP trees
to a Linear GP representation which avoids unnecessary stack
operations but enables exactly the same GP trees to be evaluated
providing a direct comparison between the two representations. To
further reduce the pressure on the L1 cache registers were used to
represent the lowest levels of the two dimensional stack. Overall,
the two dimensional stack model for GPU based GP has provided an
1.88x performance gain over the best GPU based GP approach form
the literature. A peak rate of 55.7 billion GPop/s is achieved
using a single GPU for a classification problem and 1411 billion
GPop/s for a boolean multiplexer problem.

Reducing the pressure on the L1 cache memory has increased the
effectiveness of the two dimensional stack model. Moreover, this
demonstrates that GP is as much \emph{memory bound} as
\emph{compute bound} and high performance GP is as reliant on fast
memory resources as parallel processors. Further work could
consider improving the register based stack by considering
specific interpreters based upon the the needs of a given
candidate GP program. Furthermore, combining a wider range of
memory types in an attempt to further increase the size of the
second dimension of the stack may prove beneficial. Finally, since
reducing the amount of stack memory locations required by a given
candidate GP program improves performance, further work could be
conducted in attempting to further reduce the amount of stack
memory required by GP.

\footnotesize
\bibliographystyle{agsm}      
\bibliography{2D-GPGPU}   


\end{document}